\documentstyle[11pt]{elsart}    
\def\ltsim{\raise 2pt \hbox {$<$} \kern-1.1em \lower 4pt \hbox {$\sim$}}
\def\ltapprox{\raise 2pt \hbox {$<$} \kern-1.1em \lower 5pt \hbox {$\approx$}}
\def\gtsim{\raise 2pt \hbox {$>$} \kern-1.1em \lower 4pt \hbox {$\sim$}}
\def\gtapprox{\raise 2pt \hbox {$>$} \kern-1.1em \lower 5pt \hbox {$\approx$}}

\def\skuno{\vskip 20pt}

\begin{document}
\begin{frontmatter}
\title
{Anisotropic inverse Compton scattering from 
the trans--relativistic to the ultra--relativistic
regime and application to the radio galaxies.}
\skuno
\author{G. Brunetti \thanksref{l}}

\address{
Dip. di Astronomia, Univ. di Bologna, 
via Ranzani 1, I--40127 Bologna, Italy.}
\address{
Istituto di Radioastronomia -- CNR, via Gobetti 101, I--40129
Bologna, Italy.}

\thanks[l]{gbrunetti@astbo1.bo.cnr.it}

\begin{abstract}
The problem of the anisotropic
Inverse Compton scattering between a monochromatic 
photon beam and 
relativistic electrons is revisited
and formally solved without approximations. Solutions are given 
for the single
scattering with an electron beam and with a population
of electrons isotropically distributed, under the assumption that the energy
distribution of the relativistic particles follows a simple power law as it 
is the case in many astrophysical applications. 
Both the Thomson approximation
and the Klein-Nishina regime are considered for the
scattering of an unpolarized photon beam.
The equations are obtained without the ultra--relativistic approximation and
are compared with the ultra--relativistic solutions given in the literature.
The main characteristics of the power distribution 
and spectra of the scattered radiation
are discussed for relevant examples.
In the Thomson case for an isotropic electron population
a simple formula 
holding down to the mildly--relativistic energies 
is given.

As an application the formulae of the anisotropic 
inverse Compton scattering are used to  
predict the properties of the X and $\gamma$--ray spectra 
from the radio lobes
of strong FR II radio galaxies due
to the interaction of the relativistic electrons with the incoming photons
from the nucleus.
The dependence of the emitted power on the 
relativistic electron energy distribution and on its evolution with
time is discussed.


\end{abstract}


\end{frontmatter}

\section{Introduction}

Inverse Compton (IC) scattering is one of the most common radiative 
processes in astrophysics.
The papers of Jones (1968) and Blumenthal \& Gould (1970)
provide the basic theory, while both exact expressions and
useful approximations for interaction rates, mean scattered energies,
and dispersion about the mean energies are given by Coppi \&
Blandford (1990).
The computations reported in these papers 
have been performed by supposing an isotropic 
distribution of the incident photon momenta.
However, there are astrophysical situations in which the 
photon momenta
are anisotropically distributed so that the standard equations
are not sufficient.

From a theoretical standpoint, 
Baylis et al.(1967) showed that the Compton  
spectrum (emitted over the
total solid angle) from a beam of ultra--relativistic 
electron population ($N_e(\gamma)\propto \gamma^{-\delta}$)
in the Thomson approximation is independent on the incoming
photon distribution being for all considered cases 
$\propto \epsilon^{-(\delta-1)/2}$.
The anisotropic inverse Compton (AIC) 
equations describing the scattering between a photon beam and
isotropic relativistic electrons
in the Thomson approximation 
were obtained by Bonometto et al.(1970) for a power law energy
distribution of ultra--relativistic electrons.
Aharonian \& Atoyan (1981) obtained the general {\it redistribution function} 
describing the scattering probability into a given direction of a 
monochromatic incoming
photon beam scattered by isotropic monochromatic relativistic electrons.
Furthermore, 
by integrating the {\it redistribution function} over a population
of ultra--relativistic electrons with a power law energy distribution
($N_e(\gamma)\propto \gamma^{-\delta}$) 
they obtained the analytical expression
of the spectrum (in terms of hypergeometric series) and showed that 
it presents the same shape of that derived from
an isotropic illumination
(i.e. $\propto \epsilon^{-(\delta-1)/2}$ in the Thomson
approximation and 
$\propto \epsilon^{-\delta} ln(\epsilon \epsilon_0)$
in the Klein--Nishina regime; $\epsilon_0$ being the energy of the
seed photons).   
More recently, Nagirner \& Poutanen (1993) 
have obtained the 
{\it redistribution matrix} for Compton scattering of polarized
incoming photon beams and,
by integrating the 
{\it redistribution function} over a power law energy
spectrum of ultra--relativistic electrons
in the Thomson approximation, 
they showed the equivalence between their results and those 
of Bonometto et al.(1970).

All the formulae of astrophysical interest reported in these papers
are obtained by making use of the ultra--relativistic limit.
As a consequence, 
they are not useful to the detailed study of astrophysical
situations in which also trans--relativistic and 
mildly--relativistic electrons are involved.

More recently Brunetti et al.(1997) have treated the 
AIC Thomson scattering of a photon beam by an
isotropic electron distribution
not restricted to the ultra--relativistic limit.

The AIC
has been considered in several specific astrophysical situations.
The Compton losses 
suffered by electrons moving in a bath of anisotropic
seed photons was studied both in the Thomson approximation and in the
Klein--Nishina case and was used to produce
optical and gamma--ray emission from 
the pulsars (see Morini 1981, 83 and reference therein).
With reference to the hard X--ray properties of compact sources
(AGNs and galactic black hole candidates), Ghisellini et al. (1991) have
derived (Thomson approximation) the angular distribution
of the power and the spectrum of the hard IC radiation emitted by 
a relativistic electron located at the
center of an hemispherical bowl, or cap, uniformly radiating soft
X--rays (the emission region was assumed to be optically thin and the 
AIC scattering is taken to be isotropic in the electron rest frame).
Dermer \& Schlickeiser (1993) and B\"{o}ttcher et al.
(1997) have derived approximate solutions describing the AIC scattering
of incoming photons from an accretion disc with the ultra--relativistic
electrons in the relativistic jets of AGNs.
 
Recently, Moskalenko \& Strong (1999) have discussed the effect 
due to the anisotropic distribution of the incoming photons
on the gamma--ray emission from 
interstellar photons IC scattered by cosmic--ray electrons
in the Galaxy.
This has been achieved by numerically integrating 
a {\it redistribution function},
in the Klein--Nishina regime, that was obtained
by considering
the scattered photons as unidirectionally emitted in the direction
of the electrons.
However, this approximation is strictly valid only in the
ultra--relativistic case, which is the one of interest 
in the mentioned paper.
 
The inverse Compton scattering of collimated 
streaming particles with a surrounding
bath of soft photons emitted by the disc--torus 
would produce high energy photons 
in AGNs and/or galactic X--ray binaries
(Bednarek \& Kirk 1995, Protheroe et al. 1992).
Furthermore the same mechanism has been used to model the
gamma ray emission in the MeV blazar objects (Bednarek 1998).
In order to better describe the emission pattern
of all these astrophysical situations 
general exact analytical expressions
would be useful.

The aim of
the present paper is to investigate the AIC single scattering
in the general case: from the trans--relativistic to the 
ultra--relativistic limit, both in the Klein--Nishina case and
in the Thomson approximation. General formulae of
astrophysical interest are derived.
Specifically, we will consider the scattering of a photon beam
with an electron beam (Sect.3) and with an isotropic distribution
of electron momenta (Sect.4). In Sect.2 we define the basic
geometry of the scattering and obtain general relationships
from the scattering between a monochromatic photon beam and
an electron.
In Sect.4 we also compare our results with those of
Bonometto et al.(1970) and Aharonian \& Atonyan (1981)
for the ultra--relativistic AIC and with the isotropic 
ultra--relativistic inverse Compton 
case (Blumenthal \& Gould 1970).

As an application, in Sect.5  
we use our general AIC equations of Sect.4 to study the
X--to--gamma ray spectrum of the powerful radio galaxies 
deriving from the AIC scattering of the nuclear IR--to--UV photons 
from a hidden quasar by
the relativistic electrons in the radio lobes.
In particular we argue that the detection of such 
emission with the new
generation X--ray telescopes (Chandra, XMM) can provide
important information about the magnetic field strength, 
the energy density, distribution 
and aging of the relativistic particles in the radio galaxies.

\section{IC scattering between a photon
beam and a single electron}

We assume a scattering
geometry such that the momenta of the incoming and 
scattered photon and of the electron are described
by the four-vectors (Fig.1):

\begin{equation}
P= {{\epsilon}\over c} (1,0,0,1) 
\end{equation}

\begin{equation}
P_1 = {{\epsilon_1}\over c} (1,k_j) 
\end{equation}

\begin{equation}
P_e = mc \gamma (1,\beta e_j) 
\end{equation}

\begin{figure}
\includegraphics{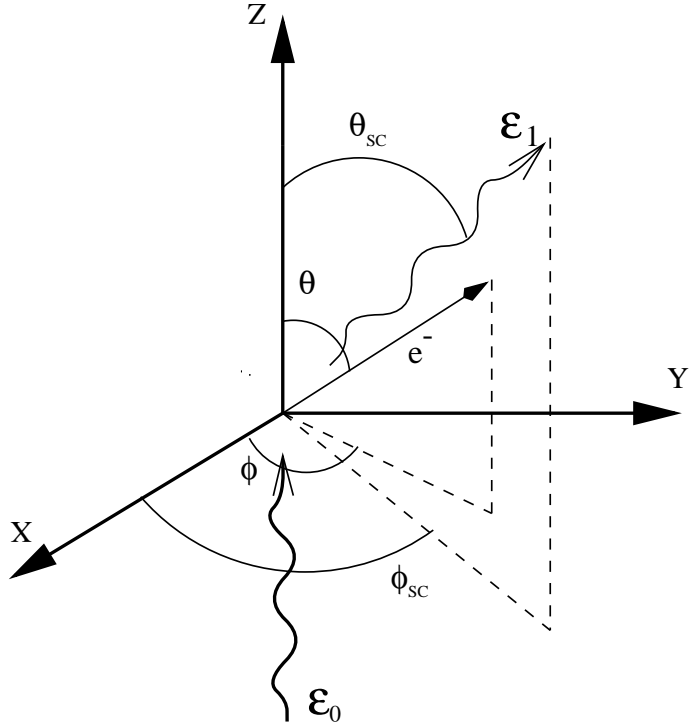}
\vspace{11 cm}
\caption{
The geometry of the scattering: 
the incoming  monochromatic (energy $\epsilon$) photon
beam propagates along the Z--axis and the photons are scattered 
to an energy $\epsilon_1$ in the
direction ($\theta_{SC},\phi_{SC}$) by electrons with momenta in the
($\theta,\phi$) direction.}
\end{figure}

$\epsilon$ and $\epsilon_1$ being the incoming and scattered 
photon energies, 
$k_j,\,e_j$ ($j=1,2,3$) the unit vector components in 
spherical coordinates, $m$ the electron
mass and $\beta=v/c$ the electron velocity.

With these assumptions the incoming photon density is given by:

\begin{equation}
n(\epsilon,\Omega_{ph}) d\epsilon d^2\Omega_{ph}=
{n\over{2\pi}} \delta(\epsilon - \epsilon_0) \delta(cos \theta_{ph}
-1) d\epsilon d^2\Omega_{ph}
\end{equation}

$\epsilon_0$ being the assumed incoming photon energy, $\delta$
the {\it Dirac--function}.

It is well known that in the Compton scattering
the tetra--vector components of the incident and scattered
photons in the lab frame and in the electron frame (primed) are
related by (Rybicki \& Lightman 1979, Pozdnyakov et al. 1983):

\begin{equation}
\epsilon_1= \gamma (1+\beta e_j^{\prime}k_j^{\prime}) \epsilon_1^{\prime} 
\end{equation}

\begin{equation}
\epsilon^{\prime}= \gamma (1-\beta e_3) \epsilon 
\end{equation}

\begin{equation}
\epsilon_1^{\prime}=\epsilon^{\prime} (1+\epsilon^{\prime}
(1-cos \theta^{\prime}_{SC})/mc^2)^{-1} 
\end{equation}

where $\gamma$ is the Lorentz factor of the electron.

In general
the inverse Compton emissivity in the lab frame
may be obtained by the manifold integral (Blumenthal \& Gould 1970):

\begin{eqnarray}
j(\Omega_{SC}, \epsilon_1)=
\int
d\epsilon^{\prime} d^2\Omega^{\prime}_{ph} d^2\Omega_e d\gamma 
{{d^3n^{\prime}(\epsilon^{\prime},\Omega^{\prime}_{ph};\epsilon,
\Omega_{ph})}\over
{d\epsilon^{\prime} d^2\Omega^{\prime}_{ph} }} 
{{d^2\Omega^{\prime}_{SC} 
d\epsilon^{\prime}_1 dt^{\prime}}\over{d^2\Omega_{SC} d\epsilon_1 dt}}
\nonumber\\ \cdot
{{d^3\sigma}\over{d^2\Omega^{\prime}_{SC}
d\epsilon^{\prime}_1}}
\epsilon_1 c \, N_e(\gamma,\Omega_e)
\end{eqnarray}

where $d^3n^{\prime}$, the photon number density in the 
electron frame, can be easily derived from the relativistic
invariant $dn/\epsilon$ (Blumenthal \& Gould 1970). 
For a monochromatic ($\epsilon_0$)
photon beam moving along the Z--axis (Eq.4) it is:

\begin{equation}
{{d^3n^{\prime}
(\epsilon^{\prime},\Omega^{\prime}_{ph};\epsilon,
\Omega_{ph})
}\over
{d\epsilon^{\prime} d^2\Omega^{\prime}_{ph} }}=
{{n}\over{2\pi}} {{\epsilon^{\prime}}\over {\epsilon}}
{{d^2\Omega_{ph} d\epsilon}\over{d^2\Omega^{\prime}_{ph} 
d\epsilon^{\prime}}} \delta(cos \theta_{ph}- 1)
\delta(\epsilon-\epsilon_0)
\end{equation}

$d\sigma$ is
the differential Compton cross section, in our case 
the unpolarized Klein--Nishina cross section (Berestetskii et al. 1982):

\begin{eqnarray}
{{d^3\sigma}\over{d^2\Omega^{\prime}_{SC}
d\epsilon^{\prime}_1}}=
{{r_0^2}\over 2} 
({{\epsilon^{\prime}_1}\over{\epsilon^{\prime}}}
)^2 ({{\epsilon^{\prime}_1}\over{\epsilon^{\prime}}}+
{{\epsilon^{\prime}}\over{\epsilon^{\prime}_1}}-1 +cos^2\theta^{\prime}_{SC}
) \cdot \nonumber\\
\delta \left(\epsilon^{\prime}_1- {{ \epsilon^{\prime}}\over
{1+\epsilon^{\prime}(1-cos\theta^{\prime}_{SC})/mc^2}} \right)
\end{eqnarray}

$N_e(\gamma,\Omega_e)$ is the electron number per energy and solid angle.
In this case it is formally written as:

\begin{equation}
N_e(\gamma,\Omega_e)d\gamma d^2\Omega_e=
\delta^2(\Omega_e-\Omega^0_e) \delta(\gamma-\gamma_0)
d\gamma d^2\Omega_e
\end{equation}

and $cos \theta^{\prime}_{SC}$, the scattering angle in the electron
frame, is derived by a Lorentz transformation of Eqs.(1--2) along
the electron direction :

\begin{equation}
cos\theta^{\prime}_{SC}=
1+ {{k_3-1}\over{\gamma^2 (1-\beta e_3)(1-\beta e_j k_j)}}
\end{equation}
 
The velocity of the electron
that scatters a photon from energy $\epsilon$ to $\epsilon_1$ is derived
by combining Eqs.(5--7) and making use of
the relativistic transformation of the angles
($\beta e_jk_j \neq 1$):

\begin{eqnarray}
\beta = \{(e_jk_j- \epsilon e_3/ \epsilon_1)^2+
({{1-k_3}\over{mc^2}} \epsilon)^2\}^{-1}
\{ 
(1-{{\epsilon}\over{\epsilon_1}})(e_jk_j
-{{\epsilon e_3}\over{\epsilon_1}}) +\nonumber\\
{{1-k_3}\over{mc^2}} \epsilon
\{ (e_jk_j- {{\epsilon e_3}\over { \epsilon_1}})^2+
({{1-k_3}\over{mc^2}} \epsilon)^2-
(1-{{\epsilon}\over{\epsilon_1}})^2 \}^{1\over 2} \}
\end{eqnarray}

that in the Thomson approximation 
(i.e. $\gamma \epsilon (1-k_3) << mc^2$) becomes

\begin{equation}
\beta \simeq
{{ \epsilon_1 - \epsilon}\over{\epsilon_1 e_jk_j -\epsilon e_3}}
\end{equation}

The Compton emissivity is readily obtained by Eqs.(8--12) and  
(5--7) and well known relativistic
transformations:

\begin{eqnarray}
j(\Omega_{SC},\epsilon_1)=
{{r_0^2 c n}\over{2\gamma_0^2}} {{\epsilon_1}\over{L_1}}
\delta(\epsilon-\epsilon_0) 
\{
(1+{{\epsilon_1}\over{mc^2}} 
{{k_3-1}\over{\gamma_0 L}} )^{-1}
+[1+ {{k_3-1}\over{\gamma_0^2 L L_1}}]^2 \nonumber\\  
+{{\epsilon_1}\over{mc^2}} 
{{k_3-1}\over{\gamma_0 L}} \}
\end{eqnarray}

where $L=1-\beta_0 e_3^0$, $L_1=1-\beta_0 e_j^0 k_j$,  
$\beta_0$ is given by Eq.(13) with $e_j$ and $e_3$ replaced with
the assumed electron coordinates 
($e^0_j$ and $e_3^0$ respectively) and

\begin{equation}
e_j^0k_j = {1\over{\beta_0}}
\{1- {{ \epsilon}\over{\gamma_0 mc^2}} (k_3-1) -{{\epsilon}\over
{\epsilon_1}} L \}
\end{equation}

In the Thomson approximation Eq.(15) becomes the much more simple:
\footnote{An alternative form of 
Eq.(17), but less useful for the purpose of this paper, 
has been recently obtained by
Fargion et al. (1997).}

\begin{equation}
j(\Omega_{SC},\epsilon_1) \simeq
{{r_0^2 c n}\over{2\gamma_0^2}} {{\epsilon_1}\over{L_1}}
\delta(\epsilon-\epsilon_0) 
\{2+2{{k_3-1}\over{\gamma_0^2 L L_1}} +
({{k_3-1}\over{\gamma_0^2 L L_1}})^2\}
\end{equation}

with $\beta_0$ given by Eq.(14) and

\begin{equation}
e_j^0k_j \simeq {1\over{\beta_0}}
\{1 -{{\epsilon}\over{\epsilon_1}} L \}
\end{equation}

\section{IC scattering between photon and electron beams}

As previously assumed the photons propagate along the Z--axis.
In this Section the
direction of the electron beam is defined by the coordinates $e_3\equiv
e_3^0\equiv cos\,\theta_e^0$ and $\phi\equiv \phi_e^0$.

The electron number density is formally given by:

\begin{equation}
N_e(\gamma,\Omega_e)d\gamma d^2\Omega_e=
K_e f(\gamma) \delta^2(\Omega_e -\Omega_e^0)
d\gamma d^2\Omega_e
\end{equation}

where $K_e$ and $f(\gamma)$ are the number density
and energy distribution of the electrons.

In this case given $\epsilon_0, \, \epsilon_1$,
the electron direction $e_j$ and the 
scattered photon direction $k_j$,  only one
electron energy ($\,\tilde{\gamma}\,$) satisfies the conservation laws. 
As a consequence 
the $\delta$--function appearing in Eq.(15) can be written as:

\begin{equation}
\delta(\epsilon -\epsilon_0)= \delta(\gamma - \tilde{\gamma})
\beta \gamma^3 J
\end{equation}

where  $\tilde{\gamma} = 
( 1- \tilde{\beta}^2 )^{-1/2}$
is given by Eq.(13) with $\epsilon_0$ instead of
$\epsilon$, while the Jacobian of $\beta$ with respect
to $\epsilon$ is:

\begin{eqnarray}
J = \{ (e_jk_j -\epsilon e_3 /\epsilon_1)^2
+((1-k_3)\epsilon /mc^2)^2 \}^{-1} 
\{ 2\beta [{{e_3}\over{\epsilon_1}} (e_jk_j - {{\epsilon e_3}\over 
{\epsilon_1}})
\nonumber\\ - 
({{1-k_3}\over{mc^2}} )^2 \epsilon ] + 
{{1-k_3}\over {2mc^2}} 
({{\epsilon d\Phi/d\epsilon}\over {\Phi^{1\over 2}}}
+ 2 \Phi^{1\over 2} )
-{{e_jk_j+e_3-2{{\epsilon e_3}\over{\epsilon_1}}}
\over {\epsilon_1}} \}
\end{eqnarray}

with $\Phi = ( (1-k_3) \epsilon /mc^2)^2+(e_jk_j - \epsilon e_3/\epsilon_1)^2
-(1- \epsilon/\epsilon_1)^2$.

The Compton emissivity is obtained by integrating Eq.(15) over the electron
distribution (Eq.19).
By making use of Eqs.(20--21) 
and well known relativistic
transformations, the integration yields: 

\begin{equation}
j(\Omega_{SC},\epsilon_1) =
{{c r_0^2 }\over{2}} n K_e \epsilon_1 \tilde{\beta}_0 
\tilde{\gamma}_0 f(\gamma= \tilde{\gamma}_0) 
\tilde{F}_0 \tilde{J}^0
(1- \tilde{\beta}_0 e_j^0k_j)^{-1}
\end{equation}

\begin{figure}
\includegraphics{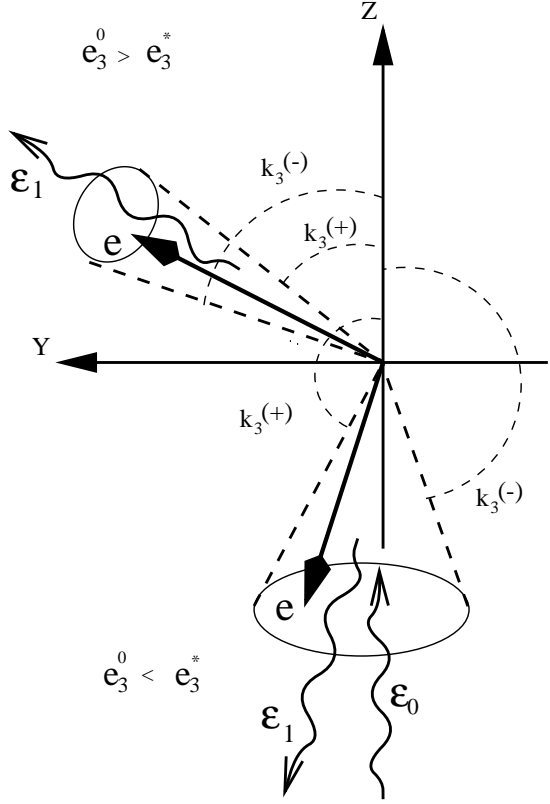}
\vspace{11 cm}
\caption{
The geometry of the scattering as seen along the X--axis.
The incoming photon beam and the scattering electron beam are shown
together with the emission cones for $e_3^0<e_3^{\ast}$ and
$e_3^0>e_3^{\ast}$.}
\end{figure}

where $\tilde{\beta}_0$ is given by Eq.(13) with the assumed
electron coordinates $e_j^0$ and the incoming photon energy $\epsilon_0$
replacing the general $e_j$ and $\epsilon$ respectively.
$\tilde{J}^0$ is given by
Eq.(21) with $e_j^0k_j$, $e_3^0$, and $\tilde{\beta}_0$
replacing
$e_jk_j$, $e_3$ and 
$\beta$, respectively, and where

\begin{eqnarray}
\tilde{F}_0 =\{1+
{{k_3 -1}\over {{\tilde{\gamma_0}^2}
(1-\tilde{\beta}_0 e_3^0) (1-\tilde{\beta}_0 e_j^0 k_j) }}  \}^2+
{{\epsilon_1}\over{mc^2}} {{k_3 -1}\over
{\tilde{\gamma}_0 (1-\tilde{\beta}_0 e_3^0) }}
\nonumber\\
+
\{ 1+ 
{{\epsilon_1}\over{mc^2}} {{k_3 -1}\over
{\tilde{\gamma}_0 (1-\tilde{\beta}_0 e_3^0) }} \}^{-1}
\end{eqnarray}

Eq.(22) can be used to calculate the emission
due to AIC scattering between a streaming population of relativistic
electrons and seed photons emitted by a small region 
(e.g., a star close to the
jet or a hot spot in the accretion disk) in AGNs or galactic 
X--ray binaries.
To describe with high accuracy much more common astrophysical
situations 
the solution can also be easily obtained by numerically integrating
over the distributions of the photon and electron momenta.

In order to describe the AIC emitted power and spectrum 
and the changes introduced by the non ultra--relativistic scattering
kinematics, we have to assume an energy distribution of the
electron population.
Let us assume a simple power law energy distribution, 
$f(\gamma) = \gamma^{-\delta}$,  so that the results of
this section can be compared with those in which the electron 
momenta are isotropically distributed (next section).
It should be born in mind, however, that there are no known 
simple mechanisms providing a power law energy distribution in accelerated
electron beams (Fermi mechanisms in general 
work with isotropic electrons momenta).

The AIC emissivity is simply obtained by replacing 
$f(\gamma= \tilde{\gamma}_0)$ with 
${\tilde{\gamma}_0}^{-\delta}$ in Eq.(22).

It can be noticed that the emitted power vanishes when $\tilde{\gamma_0}$
goes to infinity ($\delta>1$).
Electrons of increasing energy are required to scatter
the photons to a given energy $\epsilon_1$ at larger angles from
the electron beam, up to a maximum angular distance for which $\beta=1$.
Therefore photons of energy $\epsilon_1$ are scattered in the 
line of sight  
only if the angular distance of the electron beam is sufficiently
small.
Since the scattering problem is symmetric with respect to the Z--axis, it
is simpler to study the spatial distribution of the scattered radiation
by assuming that the electron beam lies on the (Y--Z) plane.
From Eq.(13), 
with $sin\, \phi_{SC}=sin\,\phi_e^0=1$, and $\epsilon=\epsilon_0$, 
one finds that the limiting condition $\tilde{\beta}_0=1$ is obtained at
two extremes in $k_3$ (Fig.2):

\begin{eqnarray}
k_3(\pm)=e_3^0[ (1-\epsilon_0(1-e_3^0)/\epsilon_1] \pm
\{ (1-(e_3^0)^2)
\nonumber\\
\cdot (1-(1-{{\epsilon_0}\over{\epsilon_1}}(1-e_3^0))^2
) \}^{1/2}
\end{eqnarray}

\begin{figure}
\includegraphics{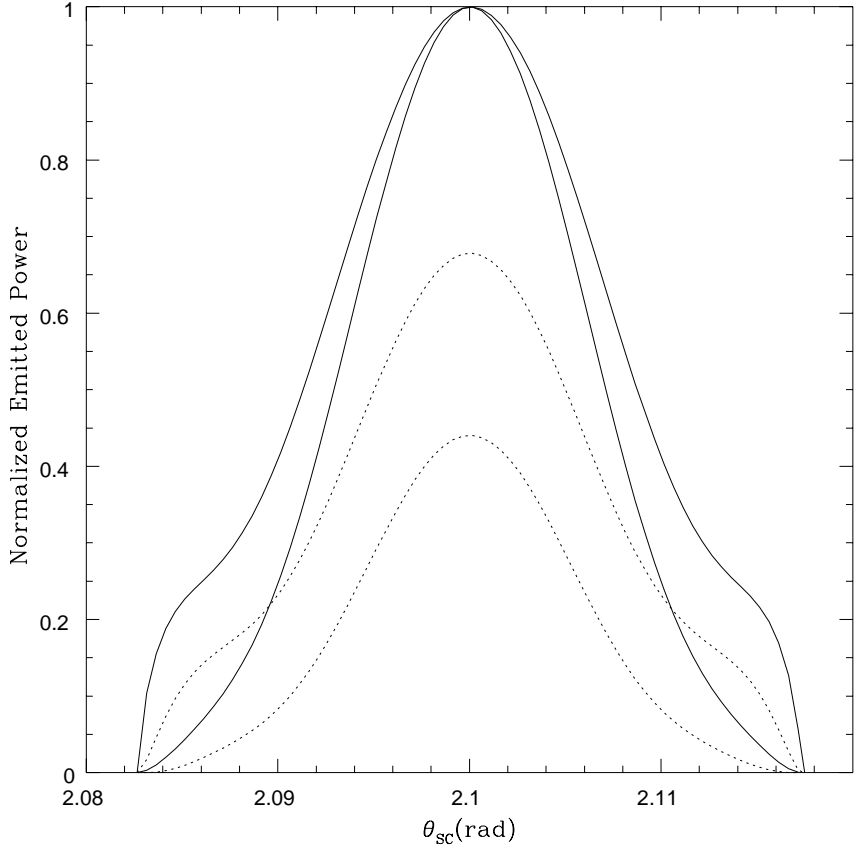}
\vspace{11 cm}
\caption{
The emitted power as a function of the scattering
angle for the IC scattering between a photon and an electron beam 
in the Thomson approximation (solid lines) 
and with the Klein--Nishina equations (dotted lines).
The curves are normalized to the maximum in the
Thomson approximation.
The electron beam is positioned at $\theta_e$(rad)=2.10 and 
$\phi_e=\pi/2$.
The upper and lower (solid and dotted) curves correspond respectively to
$\delta$= 2.5 and 5.0, while the assumed initial and scattered photon 
energies are $\epsilon_0=10^3$eV and $\epsilon_1=10$MeV.}
\end{figure}

Similarly, by releasing the condition $\sin \phi_{SC}=1$, one obtains the 
extremes in $\phi_{SC}$ by imposing the condition $k_3(+)=k_3(-)$.
If $\theta_e^0$ is large enough, the emission cone may comprise the 
Z--axis.
The angle of the electron velocity
at which this happens is derived from Eq.(24) by setting
$k_3(-)=-1$:

\begin{equation}
e^*_3 = {{-1 + \epsilon_0/\epsilon_1 }\over { 1+\epsilon_0/\epsilon_1 }}
\end{equation}

For $e_3^0\geq e_3^*$ the scattered directions on the (Y--Z) plane
are comprised in the interval
$k_3(+)<k_3<k_3(-)$ and $\phi_{SC} = \pi/2$, while for $e_3^0 < e_3^*$ 
the scattered directions are bounded by
$-1<k_3<k_3(+)$ and $\phi_{SC}=\pi/2$ and
$-1<k_3<k_3(-)$ and $\phi_{SC}=3\pi/2$ (Fig.2).

\begin{figure}
\includegraphics{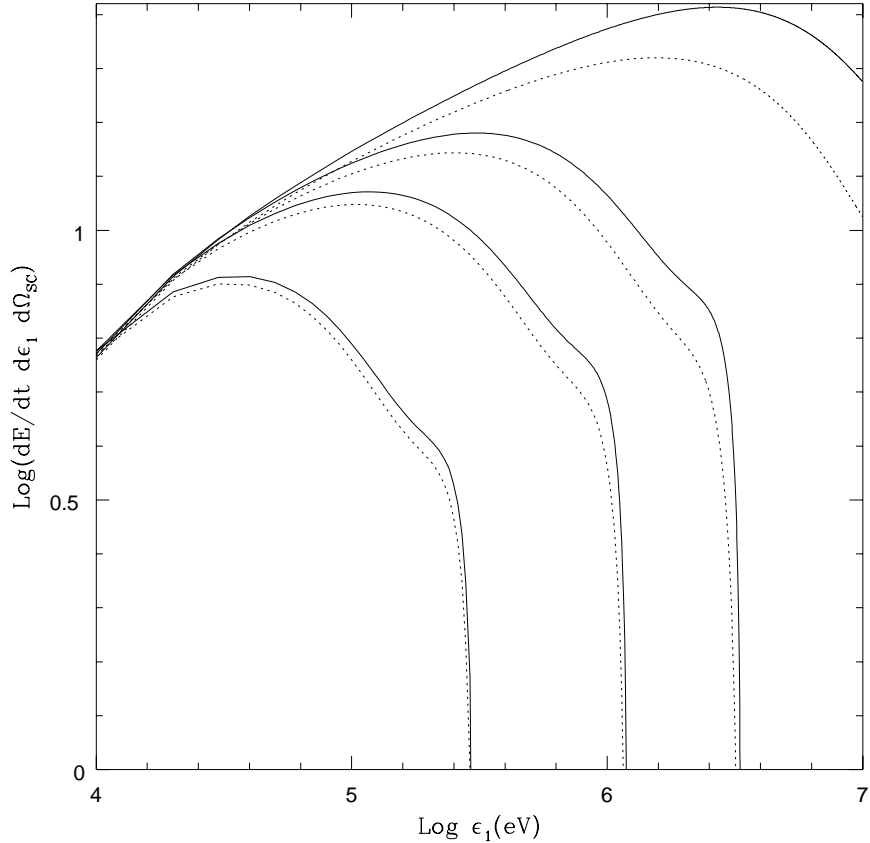}
\vspace{11 cm}
\caption{
The 
calculated spectrum from  IC scattering between a photon and an 
electron beam is shown in arbitrary units
for different scattering angles both in the
Thomson approximation (solid line) and in the 
Klein--Nishina scattering (dotted line).
We have assumed $\delta$=2.5 and an incoming photon
energy $\epsilon_0=10^3$ eV.
The assumed scattering angles from the bottom to the
top of the diagram are: $\theta_{SC}$ (rad)= 2.20, 2.15, 2.13, 2.11 
with the electron beam positioned at $\theta_e^0$(rad)=2.10 and 
$\phi_e^0=\pi/2$.}
\end{figure}

In the Thomson limit Eq.(22) becomes:

\begin{eqnarray}
j(\Omega_{SC},\epsilon_1)\simeq
{{c r_0^2 }\over{2}} n K_e {{\epsilon_1^2 (k_3-1)^{2}}\over
{\epsilon_0\, (\epsilon_1 - \epsilon_0) }}
 (1-{1\over{\tilde{\gamma_0}^2}})
\tilde{\gamma_0}^{-\delta+1}
\{
{1\over{(k_3-1)^2}} + 
\nonumber\\( {1\over{k_3-1}} 
+ {{ (\epsilon_1-\epsilon_0)^2}
\over{\epsilon_0\epsilon_1(e^0_jk_j-e^0_3)^2}}
{1\over{\tilde{\gamma_0}^2}}
)^2 \} 
\end{eqnarray}

where $\tilde{\gamma_0}$ and $\tilde{\beta_0}$ are given by Eq.(14)
with $e_j$ replaced by the assumed electron beam coordinates $e^0_j$
and with $\epsilon$ replaced by $\epsilon_0$.
Furthermore 
it can be readily
shown that in the Thomson approximation the emitted pattern
has the same boundaries of the Klein--Nishina case. 

An example of
the emitted power as a function of the scattering angle, 
both in the Thomson approximation
and in the Klein--Nishina case,  
is represented in 
Fig.3 for two different values of the electron spectral index.
Relatively more power is channeled at larger scattering
angles with decreasing $\delta$.
The broadening in the emitted power distribution in the case of low values
of $\delta$ is due to the contribution of the 
high energy electrons that emit along relatively large 
values of $\theta^{\prime}_{SC}$.

Due to the Compton recoil the energies of the electrons that are required
to scatter incoming photons ($\epsilon_0$) up to an energy $\epsilon_1$
in the Klein--Nishina regime are larger than those in the
Thomson approximation.
As a consequence the emitted power in the Klein--Nishina
regime is more depressed in the case of a steeper energy distribution
of the electrons (as it is $\delta=5$ in Fig.3).

The emitted power per unit frequency and solid angle 
as a function of the emitted energy is represented in Fig.4 for different
scattering angles both in the Thomson approximation
and in the Klein--Nishina case.
The very sharp cut-off in the emitted spectrum 
is simply due to the kinematics
of the scattering and only weakly depends on the electron differential
spectrum. 
Such feature represents the portion of the emitted spectrum contributed
by the electrons with $\beta$ approaching to unity and with the emitted
photons moving along the boundaries
of the scattered cones.

In Fig.4 it is seen that 
the difference between the Thomson
approximation and the Klein--Nishina scattering becomes larger at small
scattering angles 
(with respect to the direction of the electron beam) and
increases with the energy of the scattered photons.
Actually, for each scattering configuration this 
difference follows a quite complicated 
trend being smaller both at low and
at high emitted energies and larger at intermediate energies.
The trend is due to the difference between the Thomson and the KN 
cross sections but, because of the Compton recoil, it is also sensitive
to the electron energy distribution (see Brunetti 1998
for further details).

\begin{figure}
\includegraphics{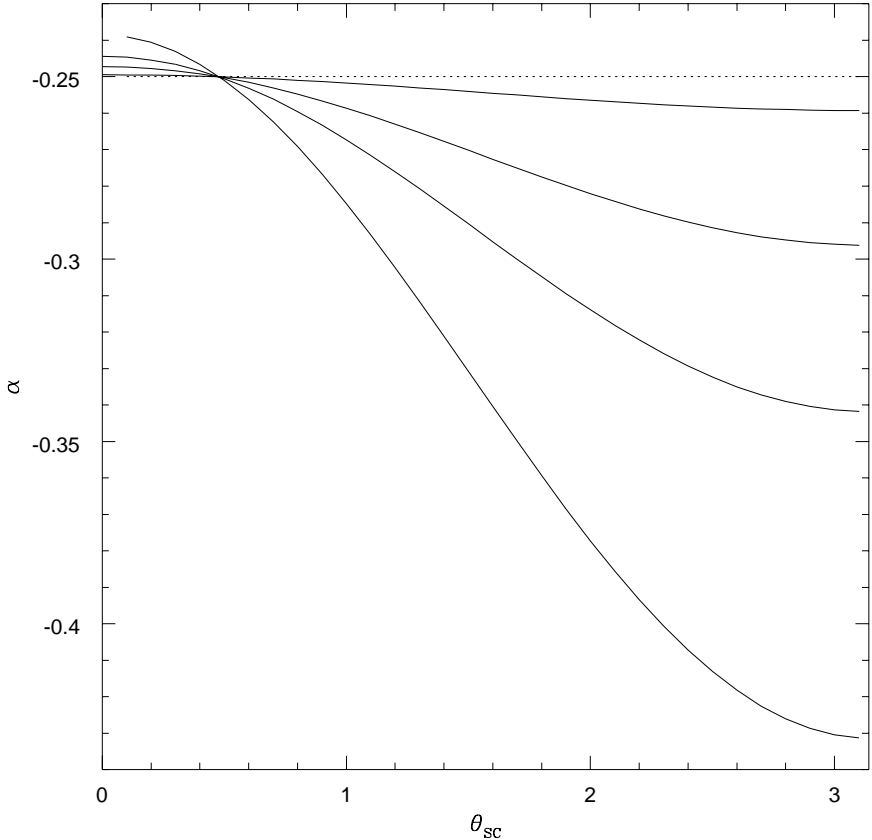}
\vspace{11 cm}
\caption{
The calculated spectral index ($j\propto \nu^{-\alpha}$)  
of the spectrum from
a scattering between a monochromatic photon beam and an electron beam
with a power law energy distribution ($\delta=2.5$).
In the scattering the photons are emitted in the direction
of the electron beam.
From the bottom of the diagram we have reported
the spectral index by assuming respectively 
$\epsilon_1/\epsilon_0$=50, 100, 200, 1000.
The dotted line is the spectral index in the
ultra--relativistic case (Eq.28).}
\end{figure}

A relevant example is the Thomson
spectrum in the direction of the
electron beam ($e_j^0k_j=1$ and $k_3=e_3^0$).
In this case
from Eq.(26) the Compton emissivity is:

\begin{eqnarray}
j(e_3^0,\epsilon_1) \simeq
{{c r_0^2 }\over{2}} n K_e {{ \epsilon_1^2(\epsilon_1-\epsilon_0)
 }\over
{\epsilon_0 (\epsilon_1- e_3^0 \epsilon_0)^{\delta+1} }}
(1-e_3^0)^{{\delta-1}\over 2}
\{1+ ( 1- 
\nonumber\\ 
( {{\epsilon_1 -\epsilon_0}\over {\epsilon_1-e_3^0 \epsilon_0}} )^2
{{ 2\epsilon_1 \epsilon_0 - (1+e_3^0)\epsilon_0^2 }\over
{\epsilon_1 \epsilon_0}} )^2 \} \nonumber\\
\cdot
\{ 2\epsilon_1 \epsilon_0 -(1+e_3^0) \epsilon_0^2 \}^{{\delta-1}\over 2}
\end{eqnarray}

\begin{figure}
\includegraphics{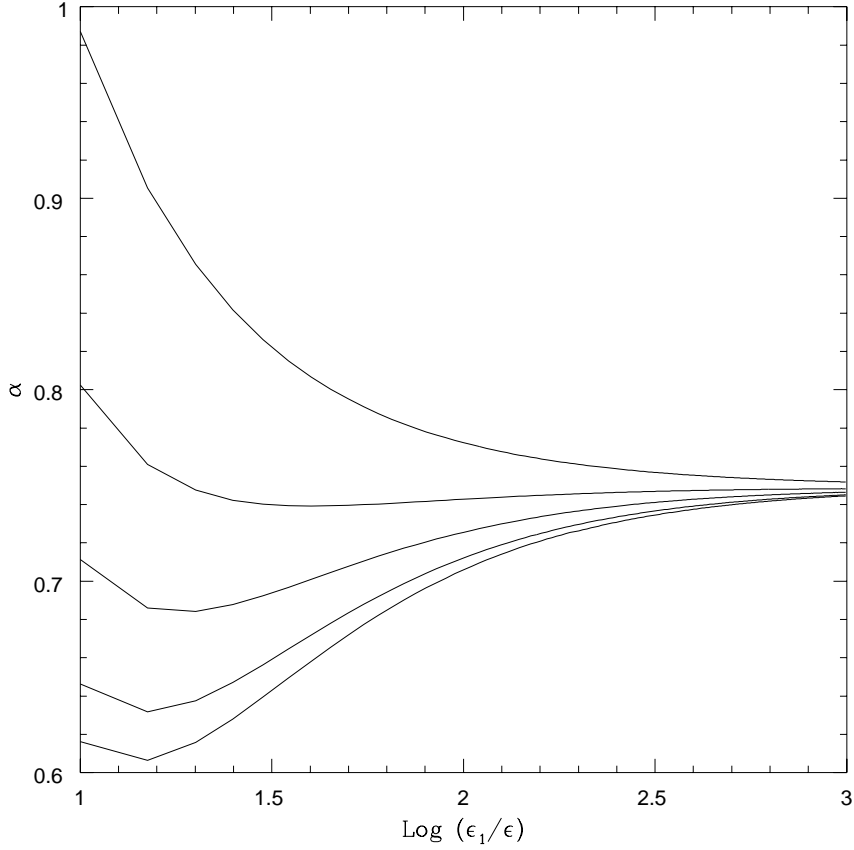}
\vspace{11 cm}
\caption{
The calculated spectral index 
of the spectrum integrated over $4\pi$ from a scattering
between a monochromatic photon beam and an electron beam
with a power law energy distribution ($\delta=2.5$) is shown as a function
of the ratio $\epsilon_1/\epsilon_0$.
From the top to the bottom of the diagram the electron beam 
direction is $\theta_e$= 0.5, 1.5, 2.0, 2.5, 3.0 (rad),
while is taken $\phi_e=\pi/2$.
The spectral index is computed with 5 eV step
($\epsilon_0$=1 eV); in the ultra--relativistic limit it
should be $\alpha=0.75$.}
\end{figure}

In the ultra--relativistic limit Eq.(27) can be expanded in series for 
$\epsilon_1>>\epsilon_0$ to get the much more simple:

\begin{equation}
j(e_3^0,\epsilon_1) \rightarrow
cr_0^2 n K_e 2^{{\delta-1}\over 2}
(1-e_3^0)^{{\delta-1}\over 2} ( {{\epsilon_1}\over
{\epsilon_0}} )^{- {{\delta-3}\over 2} }
\end{equation}

Now, as one can notice from Eqs.(5--7)
if $\theta_{SC} << 1$ then 
$\tilde{\gamma_0} >> 1$, however, 
the ultra--relativistic condition $\epsilon_1 >> \epsilon_0$ 
is not satisfied.
As a consequence although in the case $e_3^0 \simeq 1$
is $\tilde{\gamma_0} \rightarrow 1$, the
spectrum differs from the so--called ultra--relativistic case
as it can be noticed from Eqs.(27--28).

The spectrum per unit solid angle in the mildly--relativistic 
case (Eq.27) is rather complicated.
In Fig.5 its slope is shown 
as a function of the electron beam direction 
and of $\epsilon_1/\epsilon_0$; it is also compared with the 
ultra--relativistic value (Eq.28).

The power emitted by the
electron beam over $4 \pi$
can be compared with the formulae giving the emitted power
in the case of isotropic inverse Compton scattering given in the
literature (Jones 1968, Blumenthal \& Gould 1970, Rybicki \&
Lightman 1979). 
In the ultra--relativistic case the integration of Eq.(26) over
$\Omega_{SC}$ yields: 

\begin{equation}
j(\epsilon_1) \rightarrow
\pi c r_0^2 n K_e ( {{\epsilon_1}\over{\epsilon_0}} 
)^{- {{\delta-1}\over 2} } 2^{ {{\delta+3}\over 2} }
(1-e_3^0)^{ {{\delta+1}\over 2} }
{{\delta^2 +4\delta+11}\over{(\delta+1)(\delta+3)(\delta+5) }}
\end{equation}

the spectrum is a power law as in Baylis et al.(1967).
In the mildly--relativistic case the spectrum is harder for large
angles between the incoming photon and the electron beam, while it is  
softer at small angles (Fig.6).

\section{IC
scattering between a photon beam and an isotropic electron distribution}

In the case of isotropic distribution
the electron number density per energy and solid angle is:

\begin{equation}
N_e(\gamma,\Omega_e)d\gamma d^2\Omega_e=
(4\pi)^{-1} K_e f(\gamma) d\gamma d^2\Omega_e
\end{equation}

The emissivity is given by integrating Eq.(15) over the 
angular and energy
distribution of the scattering electrons.
The integral over $\Omega_e$ can be performed by following
the method given in 
Aharonian \& Atoyan (1981); one finds:

\begin{eqnarray}
j(k_3,\epsilon_1)=
{{ K_e r_0^2 \epsilon_1^2 c}\over{4 \epsilon_0^2}}
\int {{f(\gamma) }\over{\beta \gamma^2}} 
\{ {{2\epsilon_0}\over
{(\epsilon_0^2+\epsilon_1^2 -2\epsilon_0\epsilon_1 k_3)^{1/2} }}+\nonumber\\ 
( {{\epsilon_0 (1-k_3)}\over{mc^2}}
- 2{{mc^2}\over{\epsilon_1}} -
2 {{ (mc^2)^3}\over{\epsilon_0 \epsilon_1^2 (1-k_3)}} )
{\cal R}(\epsilon_0,\epsilon_1,\gamma,k_3) \nonumber\\ 
+( {{ (\gamma - \epsilon_1/mc^2 ) \epsilon_1/mc^2 +
\gamma \epsilon_0/mc^2 + \epsilon_1 \epsilon_0 k_3/(mc^2)^2
}\over{ [ (\gamma-\epsilon_1/mc^2)^2 (1-k_3)^2 +1-k_3^2 ]^{3/2}}}
\nonumber\\
+{{ (\gamma +\epsilon_0/mc^2 ) \epsilon_0/mc^2 +
\gamma \epsilon_1/mc^2 - \epsilon_1 \epsilon_0 k_3/(mc^2)^2
}\over{ [ (\gamma+\epsilon_0/mc^2)^2 (1-k_3)^2 +1-k_3^2 ]^{3/2}}})
\nonumber\\
\cdot {{1-k_3}\over{\epsilon_0\epsilon_1^2}} (mc^2)^3 \} 
n d\gamma
\end{eqnarray}

where the function

\begin{eqnarray}
{\cal R}(\epsilon_0,\epsilon_1,\gamma,k_3)=
[ (\gamma-\epsilon_1/mc^2)^2 (1-k_3)^2 +1-k_3^2 ]^{-{{1}\over2}}\nonumber\\
-[ (\gamma+\epsilon_0/mc^2)^2 (1-k_3)^2 +1-k_3^2 ]^{-{{1}\over 2}}
\end{eqnarray}

The lower limit of the 
integral Eq.(31) is given by the minimum energy of the 
electrons that can scatter an incident photon ($\epsilon_0$)
to an energy $\epsilon_1$ in the direction $\Omega_{SC}$.

The minimum velocity of the electrons  
is obtained by minimizing Eq.(13) with respect to the electron
directions ($e_j$, i.e. $\theta_e,\phi_e$).
Given that the distribution of the scattering electrons is isotropic,
it is equivalent to describe the emitted spectrum with reference
to any plane containing the incoming photon momenta (the Z--axis).
For simplicity of notation we study the 
scattering properties 
on the (Y--Z) plane, so that $k_1=0$, $k_2=(1-k_3^2)^{1/2}$, and 
the 
electrons having the lowest energy to make the scattering 
are those moving on the (Y--Z) plane i.e. with $\phi_e=\pi/2$
(i.e. $e_1=0$).

We find that Eq.(13) is minimized by electrons having
directions :

\begin{equation}
e_3(\pm)=
\pm \{ 1+ {{ k_2^2}\over{(k_3- \epsilon_0/\epsilon_1)^2}} \}^{-1/2}
\end{equation}

with $e_3(+)$ for $k_3\geq \epsilon_0/\epsilon_1$ and $e_3(-)$
for $k_3 < \epsilon_0/\epsilon_1$.
By combining Eqs.(13) and (33) one finds $\beta_{min}$ and 
$\gamma_{min}=\{1-\beta_{min}^2\}^{-1/2}$.
It should be noticed that Eq.(33) holds whatever the electron
and the photon energy may be,  i.e. both in the Thomson approximation
and in the Klein--Nishina
case, from the ultra--relativistic limit
down to the trans--relativistic case.

In the Thomson approximation, 
from Eqs.(14) and (33), we find:

\begin{equation}
\gamma_{min}\simeq
\{1+ {{ (\epsilon_1 -\epsilon_0)^2}\over
{2\epsilon_0\epsilon_1(1-k_3)}} \}^{1/2}
\end{equation}

that becomes the simplest
$\gamma_{min}\rightarrow(\epsilon_1/2\epsilon_0(1-k_3))^{1/2}$ in the
ultra--relativistic case (Bonometto et al. 1970), and
the well known 
$\gamma_{min}=1/2\sqrt{\epsilon_1/\epsilon_0}$
in the ultra--relativistic isotropic Thomson
scattering (taking $k_3=-1$).

\subsection{The Thomson approximation}

In the literature it is usually assumed
a momentum--power law 
energy distribution of the relativistic
electrons $f(\gamma) =\beta^{-1} (\beta \gamma)^{-\delta}$
since it is
that expected from Fermi acceleration mechanisms.
It deviates from a simple
energy--power law in the non ultra--relativistic case.

In this section our aim is to derive AIC equations from the 
trans--relativistic to the ultra--relativistic regime and to compare
our findings with the ultra--relativistic equations in the
literature.
As a consequence, in order to evidence the changes in the AIC emitted
spectrum and pattern introduced only by the non ultra--relativistic
scattering kinematics, in this section, we adopt 
a simple energy--power law distribution
$f(\gamma) =\gamma^{-\delta}$ down to trans--relativistic energies.
The case of a general
$f(\gamma)$ is discussed in Appendix A where 
a simple semi--analytical
solution is given.
The case of the distribution 
$f(\gamma) =\beta^{-1} (\beta \gamma)^{-\delta}$ 
is discussed in Appendix B
and is compared with the results of this section.

From Eq.(31) the integration in the Thomson case yields:

\begin{eqnarray}
j(k_3,\epsilon_1) \simeq
{{K_e r_0^2 c}\over{4}} ( {{\epsilon_1}\over{\epsilon_0}} )^2
n
\{
{{2 {\cal I}_0 }\over
{( ( {{\epsilon_1}\over{\epsilon_0}})^2 -2k_3 {{\epsilon_1}\over{\epsilon_0}}
+1)^{1\over 2} }} - \nonumber\\
2(1-k_3)^{-1}(1+{{\epsilon_0}\over{\epsilon_1}}) {\cal I}_{3/2}+\nonumber\\
(1-k_3)^{-2} 
[(1+{{\epsilon_0}\over{\epsilon_1}})(3k_3-{3\over 2}) -{3\over 2}
({{\epsilon_1}\over{\epsilon_0}}+({{\epsilon_0}\over{\epsilon_1}})^2)]
{\cal I}_{5/2}
\nonumber\\
+{5\over 2}(1-k_3)^{-2}
[3(1+{{\epsilon_0}\over{\epsilon_1}})+({{\epsilon_1}\over
{\epsilon_0}}+ ({{\epsilon_0}\over{\epsilon_1}})^2 ) ] 
{\cal I}_{7/2} \}
\end{eqnarray}

where

\begin{equation}
{\cal I}_0=
{1\over{\sqrt{\pi}}} \sum_{i=0}^{\infty}
{{\Gamma(i+{1\over 2}) }\over
{(\delta+2i+1)\Gamma(i+1)}}
\gamma_{min}^{-\{\delta+2i+1\} }
\end{equation}

\begin{eqnarray}
{\cal I}_{3/2}=
{{2}\over{\pi}} \sum_{i,m=0}^{\infty}
{{(m+{1\over 2}) \Gamma(m+{1\over 2}) \Gamma(i+{1\over 2})
 }\over{
(g_{i,m}-2) \Gamma(i+1) \Gamma(m+1) }}
\gamma_{min}^{-\{g_{i,m}-2\} } S^m_{k_3}
\end{eqnarray}

\begin{eqnarray}
{\cal I}_{5/2}=
{4\over{ 3 \pi}}
\sum_{i,m=0}^{\infty}
{{ (m^2+2m+{3\over 4} ) \Gamma(m+{1\over 2}) \Gamma(i+{1\over 2})
 }\over{
g_{i,m} \Gamma(m+1) \Gamma(i+1) }}
\gamma_{min}^{-g_{i,m} } S^m_{k_3}
\end{eqnarray}

\begin{eqnarray}
{\cal I}_{7/2}=
{8\over{15 \pi}}
\sum_{i,m=0}^{\infty}
{{ (m^3+ {9\over 2} m^2 + {{23}\over 4} m + {{15}\over 8} )
\Gamma(m+{1\over 2}) \Gamma(i+{1\over 2}) 
}\over{g_{i,m} \Gamma(i+1) \Gamma(m+1) }}
\gamma_{min}^{-g_{i,m} } S^m_{k_3}
\end{eqnarray}

where $S_{k_3} \equiv (k_3^2-1)/(1-k_3)^2$, 
$g_{i,m} \equiv \delta+2(i+m)+5$, and $\gamma_{min}$ is given
by Eq.(34).
\footnote{
It would also be noticed that the ratios $\Gamma(i[m]+1/2)/
\Gamma(i[m]+1)$ in Eqs.(37--39) are also equivalent to
$4^{i[m]} B(i[m]+1/2, i[m]+1/2) / \pi^{1/2}$; $B$ being the
{\it Beta function}.}

\begin{figure}
\includegraphics{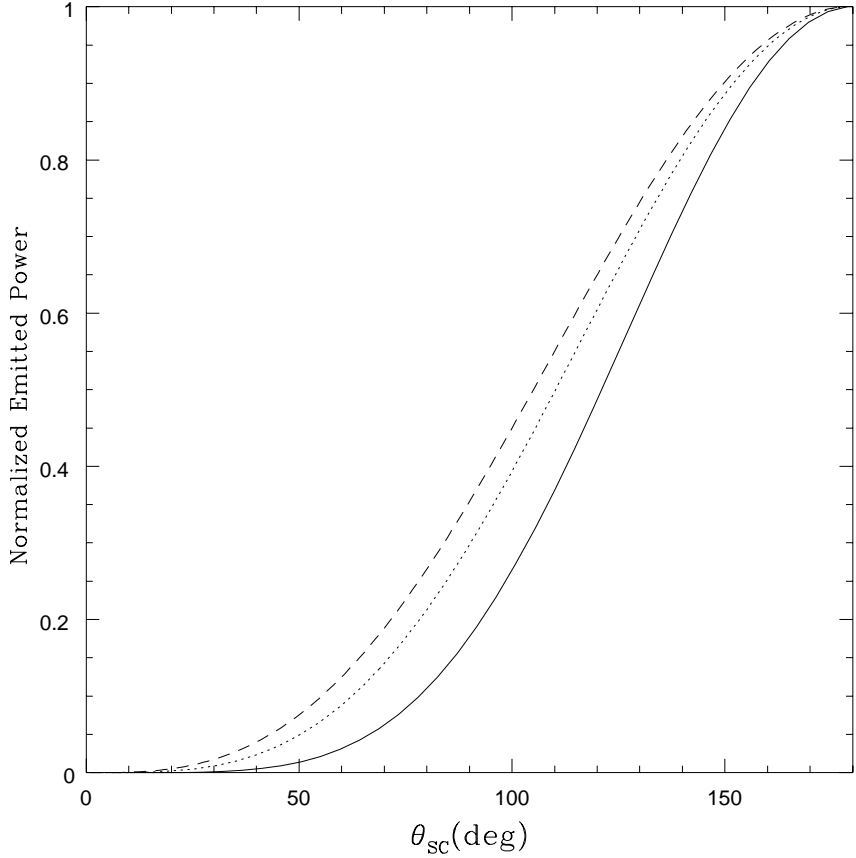}
\vspace{11 cm}
\caption{The normalized emitted power as a function of the scattering 
angle for different values of $\delta$: 2.0 (dashed line), 
2.5 (dotted line), 4.0 (solid line).
In the computation 
it has been assumed $\epsilon_0$=0.1 eV and $\epsilon_1$=1 keV
i.e. Thomson ultra--relativistic case.
Relatively
more power is emitted al large $\theta_{SC}$ with increasing $\delta$.}
\end{figure}

A solution in a simple closed form for Eqs.(35--39) 
that holds down to the mildly--relativistic case 
(for $\epsilon_1/\epsilon_0 \geq 15-20$ with $< 5\%$ approximation)
is:

\begin{equation}
j(k_3,\epsilon_1) \simeq
K_e r_0^2 c n (1-k_3)^{ {{\delta+1}\over 2} } \left( 
{{\epsilon_1}\over{2\epsilon_0}} \right)^{ -{{\delta-1}\over 2} }
\{A_1(\delta) + {{\epsilon_0}\over{\epsilon_1}} A_2(\delta,k_3) \}
\end{equation}

where

\begin{equation}
A_1(\delta) = {{ \delta^2 +4\delta +11}\over{(\delta+1)(\delta+3)
(\delta+5) }}
\end{equation}

and

\begin{equation}
A_2(\delta,k_3)=
{{ (\delta^4 + 8\delta^3 +44 \delta^2 +128\delta +139)k_3
-3 \delta^3 +\delta^2 +35\delta+31}\over{(\delta+1)(\delta+3)
(\delta+5)(\delta+7) }}
\end{equation}

This very simple formula 
could be very useful in astrophysical problems 
where non ultra--relativistic
electrons are involved in the scattering process.

\begin{figure}
\includegraphics{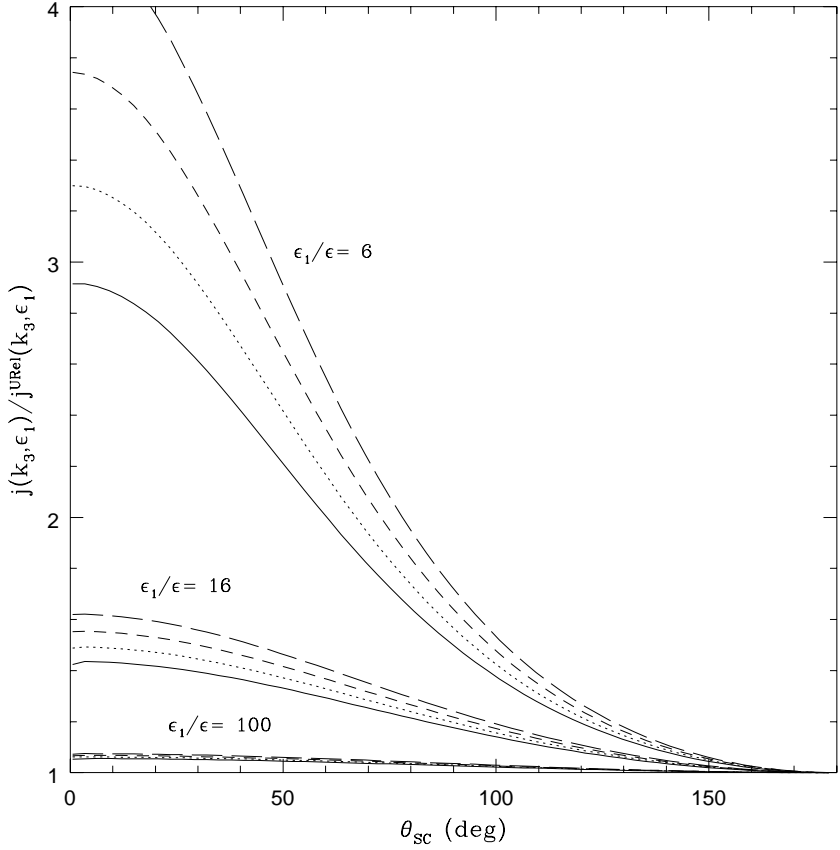}
\vspace{11 cm}
\caption{
The ratios between the general AIC and the ultra--relativistic
($j^{URel}$) emitted power as a function of the scattering
angle ($\theta_{SC}$) are given for $\epsilon_1/\epsilon$=
6, 16, 100.
In the calculation it is assumed $f(\gamma)=\gamma^{-\delta}$
with $\delta$= 2 (solid lines), 2.5 (dotted lines), 3 (small--dashed
lines), and 3.5 (large--dashed lines).
All the calculated patterns are normalized at $\theta_{SC}
=180^o$.
One finds that the percentage of power emitted at small
$\theta_{SC}$ increases in the mildly and trans--relativistic
case.}
\label{sample-figure}
\end{figure}

\begin{figure}
\includegraphics{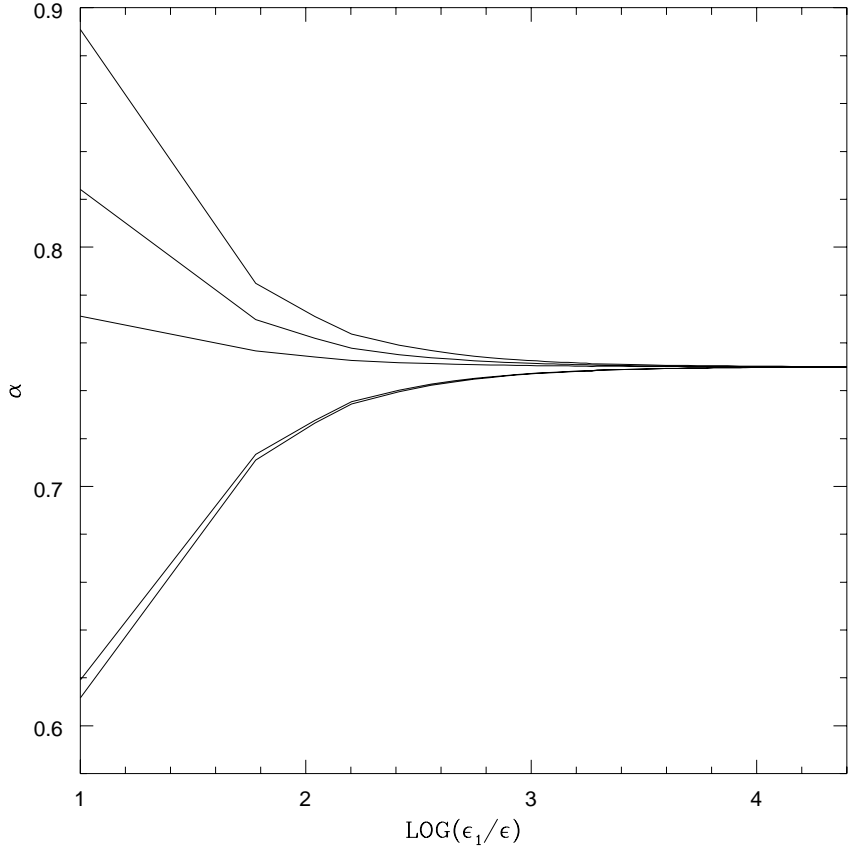}
\vspace{11 cm}
\caption{
The calculated spectral index ($j\propto \nu^{-\alpha}$)
computed with a 5 eV step with
$\epsilon_0$=0.1 eV) in the
Thomson regime as a function of $\epsilon_1/\epsilon_0$
for several values of the scattering angle.
Starting 
from the top the scattering angles are: 
$\theta_{SC}$(rad)= 0.8, 1.2, 1.5, 2.8, 3.1.
We have assumed $\delta$=2.5 and $\epsilon_0$=0.1 eV.
At lower energies and 
large scattering angles the spectrum is flatter than  
in the highly relativistic  
isotropic case ($\alpha=0.75$), 
while it is steeper at 
small scattering angles.
The differences become vanishing with increasing the emitting frequency.}
\label{sample-figure}
\end{figure}
 
By considering Eqs.(36-39) in the ultra--relativistic limit
$\epsilon_1>>\epsilon_0$, after simple algebraic manipulation
from Eq.(35) (or simply from Eqs.(40--42)) we find the emissivity in the 
ultra--relativistic case:

\begin{equation}
j(k_3,\epsilon_1)\rightarrow
K_e r_0^2 c n 
{{(1-k_3)^{ {{\delta+1}\over 2} } 
( \delta^2 +4\delta +11)}\over{(\delta+1)(\delta+3)(\delta+5)}}
({{\epsilon_1}\over{2\epsilon_0}})^{- {{\delta-1}\over 2} }
\end{equation}

that is equivalent to the result given in Bonometto et al.(1970).
Furthermore by integrating Eq.(43) over $\Omega_{e}$
(it should be noticed that 
due to the isotropic distribution of the electron
momenta it is equivalent to integrate over $\Omega_{SC}$) 
one finds the well known solution 
in the isotropic ultra--relativistic 
Thomson scattering case (Blumenthal \& Gould 1970). 

The emitted power is obtained by Eq.(35) in the general case or by
the much more simple
Eq.(43) in the ultra--relativistic case; 
in the ultra--relativistic case, it is shown in Fig.7. 
As expected the scattered power has a maximum 
at $\theta_{SC}=\pi$ and goes to zero for small scattering angles.
The electrons that are mainly responsible 
for the scattered power at a given $\epsilon_1$ are those with energies
close to the minimum energy that increases with the scattering angle.
It follows that by increasing $\delta$ there are relatively fewer
electrons at higher energies and, consequently, the emitted power 
goes to zero more rapidly toward smaller scattering angles.

In order to compare our general results with the AIC formulae 
in the literature we give  in Fig.(8) the ratios between
the AIC emission (Eq.35) and that of the 
ultra--relativistic case
(Eq.43) as a function of the scattering angle
for different electron energy distributions and energies; 
the differences are $> 30 \%$ also in the mildly--relativistic 
case (i.e. $\epsilon_1/\epsilon \sim 20-30$).
The main effect due to the non--relativistic kinematics is that
the percentage of radiation emitted at large scattering angles
decreases with decreasing $\epsilon_1/\epsilon$.

\begin{figure}
\includegraphics{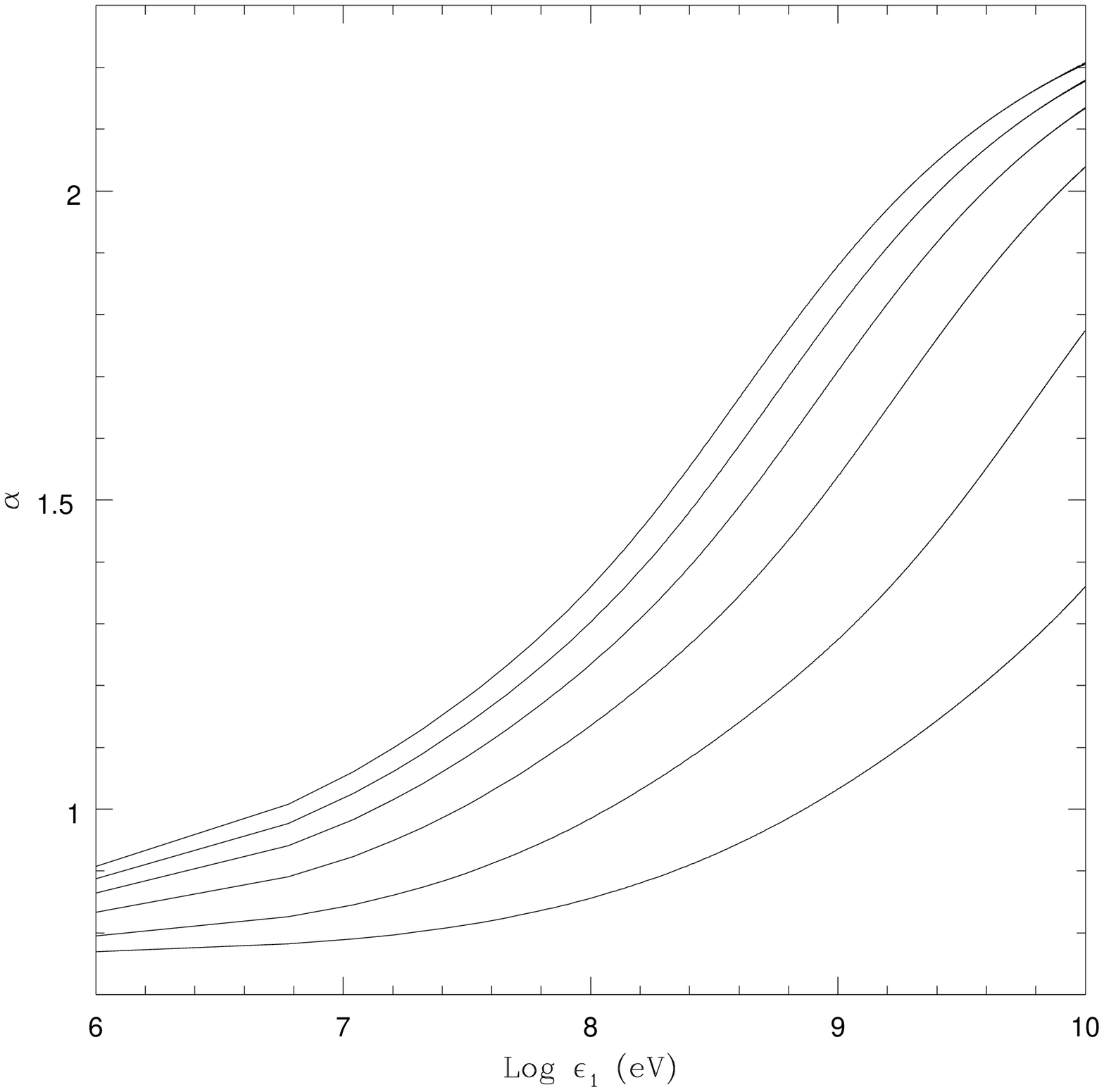}
\vspace{11 cm}
\caption{
The calculated spectral index 
(computed with 5 MeV step; $j \propto \nu^{-\alpha}$) in the Klein--Nishina
regime for an isotropic distribution of the relativistic electrons 
as a function of the emitted energy for different values
of the scattering angles.
We have assumed $\delta$=2.5 and a monochromatic incident photon beam
($\epsilon_0=10^3$ eV).
Starting from the bottom of the diagram the successive values of the
scattering angles are: 
$\theta_{SC}$(rad)= 0.2, 0.5, 1.0, 1.5, 2.0, 2.8.
The Klein--Nishina spectrum steepens at higher energies with decreasing
scattering angle.}
\label{sample-figure}
\end{figure}

The emitted spectrum for different scattering
angles and $\delta$=2.5 is shown in Fig.9.
As already noticed (Eqs.35 and 42),  
at variance with the 
ultra--relativistic standard equations in the literature, 
the AIC spectral index calculated with our general equations
is a function of
the scattering angle and of the energy.
With decreasing $\epsilon_1/\epsilon$ the 
spectrum flattens at large $\theta_{SC}$ and steepens at
small $\theta_{SC}$; this causes the progressive
decrease of the percentage of radiation emitted at
large $\theta_{SC}$ (Fig.8).
The difference from the standard 
ultra--relativistic spectrum (that is $\alpha=0.75$), 
vanishes at
higher energies due to the relativistic aberration.

\subsection{The Klein--Nishina case}

It is well known that
the emitted IC
spectrum, in the case of an isotropic distribution of the seed photons
momenta, typically steepens at an energy ($\epsilon_{KN}$) 
such that  $\epsilon_{KN}^{\prime} \sim mc^2$.

\begin{figure}
\includegraphics{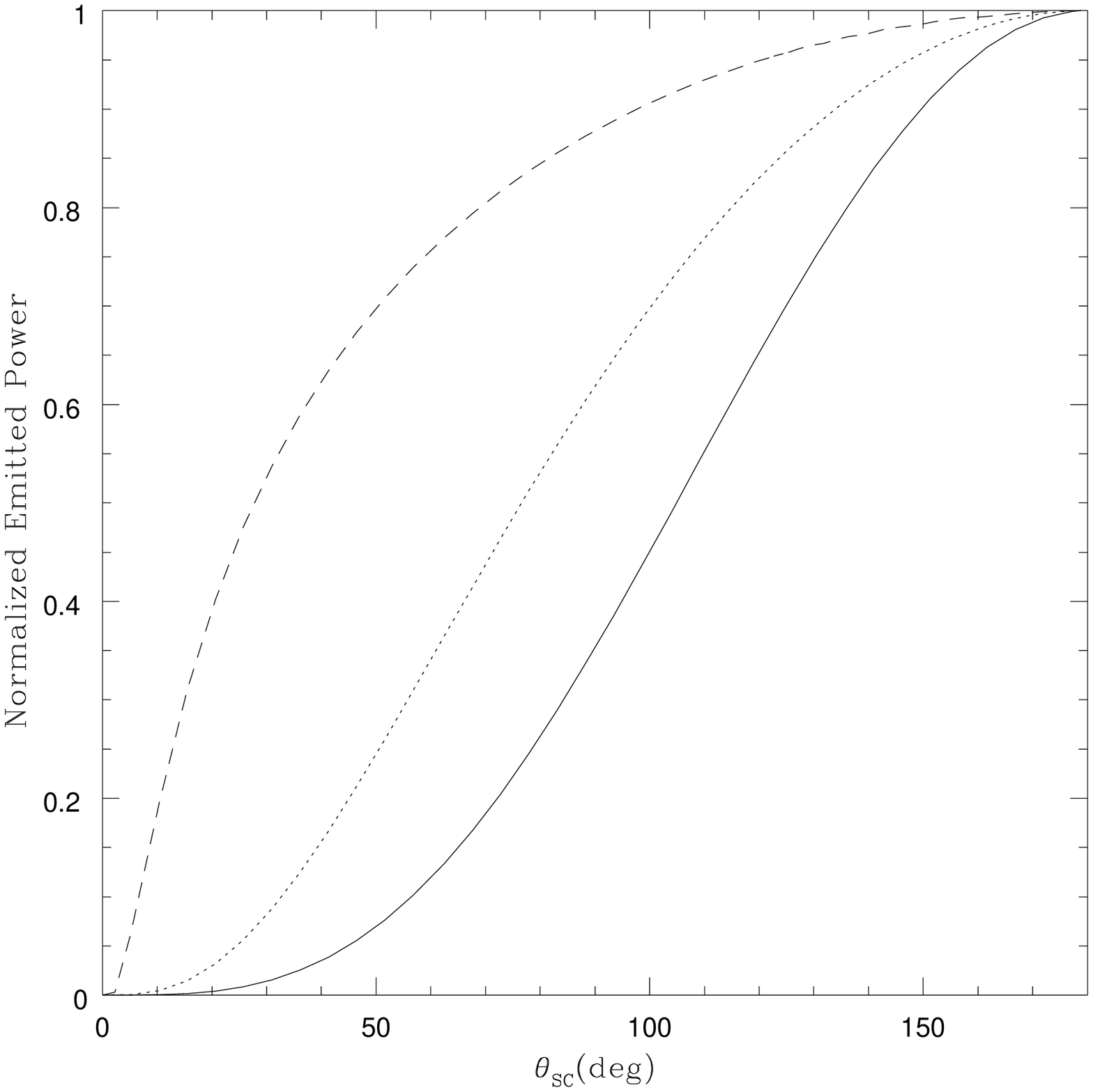}
\vspace{11 cm}
\caption{The normalized emitted power as a function of the scattering 
angle in the Klein--Nishina regime is plotted for three different values
of the emitted energy: $\epsilon_1$= 10$^7$ eV (solid line), 
10$^9$ eV (dotted line), 10$^{11}$ eV (dashed line).
We have assumed $\delta=2.5$, and an incoming photon energy $\epsilon_0$=
10$^3$ eV.
The emitted power is distributed much more isotropically (KN--phase 
isotropization) at larger emitted energies.}
\end{figure}

\begin{figure}
\includegraphics{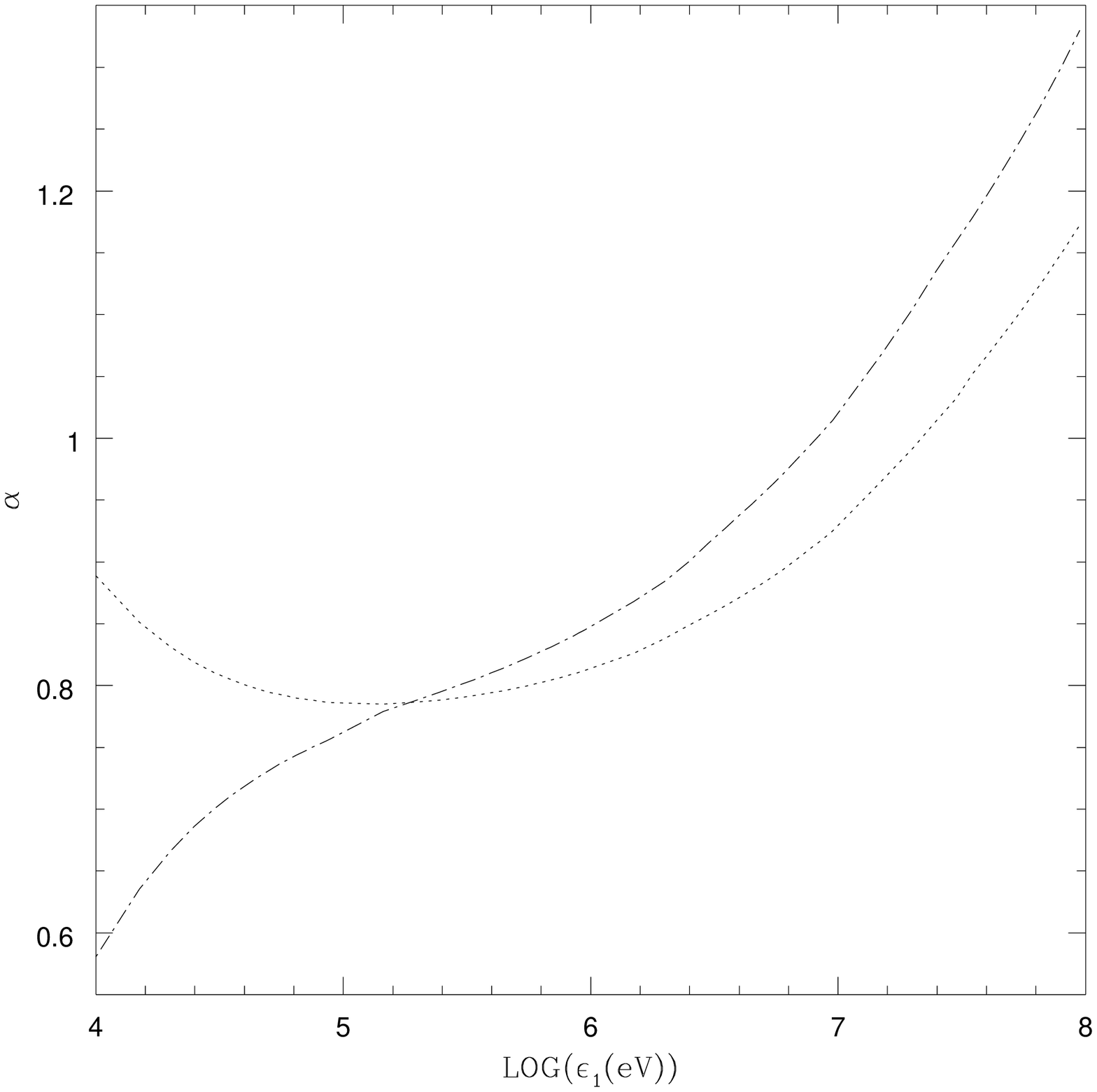}
\vspace{11 cm}
\caption{The calculated spectral index 
($j\propto \nu^{-\alpha}$ computed with a 10$^4$ eV
step) in the Klein--Nishina regime as a function of the emitted
frequency for two different values of the scattering angle:
$\theta_{SC}$= 1.2 (dotted line), 2.4 (dot-dashed line).
We have assumed $\delta$=2.5 and an incident photon energy
$\epsilon_0$ =10$^3$ eV.
In the diagram both the departure from the classical isotropic
spectral index at low photon energies and the spectral steepening 
due to the Klein--Nishina effect are shown.}
\label{sample-figure}
\end{figure}

In the case of the AIC scattering $\epsilon_{KN}$
is expected to depend on the scattered directions.
Eq.(31) 
can be used to investigate the Klein--Nishina steepening
at different scattering angles both in the ultra--relativistic and in
the general case.
This point is illustrated in Fig.10 where we have plotted the
AIC spectra for different values of $\theta_{SC}$ 
by assuming a power law energy distribution
of the scattering electrons ($f(\gamma)=\gamma^{-\delta}$, 
$\delta$=2.5 and $\epsilon_0$=1 keV).
The behaviour in Fig.10
is readily understood by keeping in mind 
the kinematics of the Compton scattering (Eqs. (5) and (7)).
One has that $\epsilon_{KN} \sim
\gamma \epsilon_{KN}^{\prime} \sim \gamma\, mc^2$  and that $\gamma$,
the typical (or minimum) energy of the  electron
required to scatter a photon ($\epsilon_0$) to $\epsilon_1$
increases for large $\theta_{SC}$.
As a consequence the KN--phase starts at smaller 
emitted energies with increasing $\theta_{SC}$. 
On the other hand, we have shown that in the Thomson approximation
the emitted power is greatly enhanced at large $\theta_{SC}$
(Fig.7).
These two effects can largely compensate each other.
This is clearly illustrated in Fig.11 where we have plotted
the emitted power as a function of the scattering angle (with
$\delta$=2.5 and $\epsilon_0$=1 keV) for several values of the
scattered photon energy $\epsilon_1$.
When the KN regime is well established (larger
$\epsilon_1$) the scattered power is much more isotropically distributed
(a sort of KN isotropization).
 
As shown by Aharonian \& Atoyan (1981), 
the shape of the 
AIC emitted spectrum in the ultra--relativistic case
is identical to the isotropic case.
Aharonian \& Atoyan (1981)
Equation can be obtained by integrating Eq.(31) over the
electron spectrum in the ultra--relativistic limit.

However, 
since the KN regime may become important at relatively low energies of
the scattering electrons, both the non ultra--relativistic
effects previously discussed (Fig.9) 
and the classical KN 
steepening may show up in the emitted spectrum.
This happens to be in the energy range that simultaneously
satisfies $\epsilon_1 \epsilon_0 (1-k_3) \geq 10^{-2} (mc^2)^2$ 
(for which the spectrum becomes sensitive to the KN effects) and
$\epsilon_1/\epsilon_0 \leq 10^{2.5}$ 
(i.e. non ultra--relativistic).
It is illustrated in Fig.12 where we have plotted 
the spectral index as a function of the scattered photon energy for
two representative values of $\theta_{SC}$ (with $\delta$=2.5 and
$\epsilon_0$=1 keV).
Our calculations show that at low energies the spectral
index is much steep in the case of the smaller $\theta_{SC}$ and
starts to converge toward the classical ultra--relativistic
limit ($\alpha$=0.75 in this example) but the progressive KN steepening
of the spectra takes over before the ultra--relativistic regime is
established.
When the KN regime is well established the spectral index
is much steep in the case of the large $\theta_{SC}$.

\section{Application to the FR II radio galaxies}

By adopting the unification
scheme linking radio loud
quasars and FR II radio galaxies (Barthel 1989), 
it has been shown that 
the AIC scattering of the nuclear
photons by the relativistic electrons in the lobes may
produce detectable X--ray fluxes 
well in excess of those calculated by the IC of the CMB 
photons (Brunetti et al.1997).

Roughly speaking, 
the inverse Compton emissivity is 
$\propto K_e L_{QSO}/R^2$, $R$ and $L_{QSO}$ being 
the distance from the hidden quasar and 
the far IR--UV quasar luminosity respectively.
By constraining $L_{QSO}$, a comparison of the
X--ray properties with the radio synchrotron emission
can provide important information on the magnetic
field strength and relativistic particle distribution
in the lobes.

In a recent ROSAT HRI observation of the FR II radio galaxy 3C 219 
Brunetti et al.(1999) have found 
extended non--thermal emission from the lobes
most likely due to AIC 
of nuclear photons by relativistic electrons.
It was found that the
average magnetic field strength of 3C 219 
is a factor $\sim$3 lower than 
that computed under equipartition assumption.

In order to illustrate the possibility of obtaining
detailed physical information from X--ray observations, 
which will be possible with the future Chandra and XMM
missions, in
this Section we apply the general AIC equations of Sect.4
to a specific model.

\subsection{Calculation of the emissivity}

We assume that the far IR--optical spectrum of a typical quasar
hidden in a powerful radio galaxy 
is represented by four power laws ($F_{\nu}^{-\alpha}$) 
with $\alpha$ = 0.2,
0.9, 1.7, and 0.6 
respectively in the intervals 100--50, 50--6, 6--0.65,
and 0.65--0.35 (as in Brunetti et al.1997). 
Since the UV photons may be important in the IC production
of the hard X and gamma--rays, in addition we assume a 
UV spectrum simply modelled by a single power law with  
$\alpha$ = 0.0 in the 0.35--0.03$\mu$m band, 
consistently with Walter \& Fink (1994) findings.

We further assume that 
the relativistic electrons are injected in the radio volume, 
approximated with a prolate ellipsoid of $100\times 50$ kpc
semiaxis, 
with an energy differential spectrum 
$N_e(\gamma)=K_e \gamma^{-\delta}$
(the calculated AIC fluxes are typically contributed by electrons by 
$\gamma> 5-10$) and
that the magnetic fields, of average constant intensity $B$, 
and particles momenta are randomly
distributed on a sufficiently small scale. 

The time evolution of the particle spectrum, injected at the hot spots
and no reacceleration, due to synchrotron and Compton losses may
be described by (Jaffe \& Perola 1973):

\begin{equation}
N_e(\gamma) \propto 
\gamma^{-\delta} (1-{{\gamma}\over{\gamma_b}})^{\delta-2};
\,\,\,\, for \,\,\, \gamma<\gamma_b
\end{equation}

being $N_e(\gamma)=0$ for $\gamma>\gamma_b$ and 
$\gamma_b$ the break energy (Kardashev 1962).
This is crucial in order to compare the predictions with detailed 
spatially resolved X--ray spectroscopy.
Deep radio observations of the hot spot regions of the radio galaxy Cyg A
have suggested the possibility  of a low energy
cut off, or a turn over in the particle distribution
(Carilli et al. 1991). 
We take into account this possibility by applying an
exponential cut off to the power law spectrum below $\gamma_c$.

\begin{figure}
\includegraphics{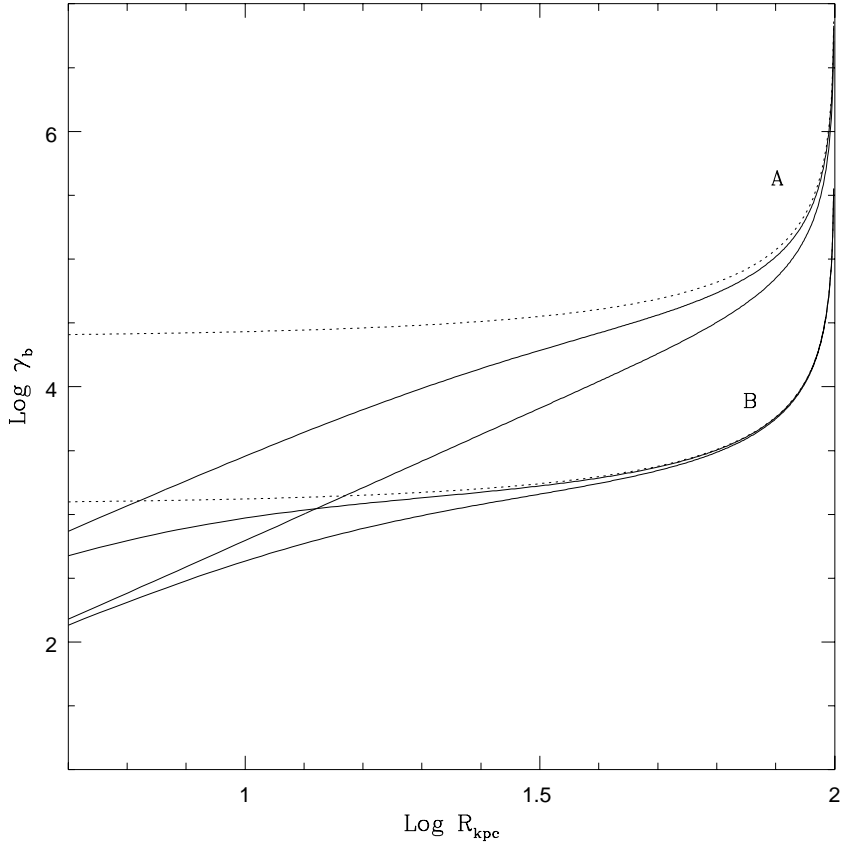}
\vspace{11 cm}
\caption{
The calculated break energy (in units of $mc^2$) is reported
as a function of the distance from the nucleus $R_{kpc}$ for different
magnetic field strengths $B$=1$\cdot 10^{-5}$ (A)
and 5$\cdot 10^{-5}$ G (B).
A distance of the hot spot from the nucleus $D$=100 kpc, 
a velocity of separation $\eta$=0.04 and a redshift $z$=0.15 have been
assumed. The luminosity of the hidden quasar 
from the bottom to the top of the diagram (solid lines) are:
$5$ and $1\cdot 10^{46}$erg s$^{-1}$.
The dotted lines represent the prediction without the hidden quasar.}
\end{figure}

In general, $\gamma_b$ is expected to be
a function of the position $R$ in the radio galaxy.
In a very simple model, in which the radio hot spots 
separate with constant
velocity ($\eta c$) and the particles remain approximatively in the same
place in which they are injected, the age 
of the relativistic particles 
is $t = (D - R)/(\eta c)$, D being the distance of the
hot spot (for which $t$=0) from the nucleus.

At variance with the previous literature, 
in addition to the standard synchrotron and Compton (with the
CMB photons) losses, we consider also
the radiative losses due to the AIC scattering 
of the nuclear photons.

With these assumptions the break energy 
of the time/spatially evolved electron population
in the radio galaxy is:

\begin{equation}
\gamma_b(R_{kpc}) =
{{5\cdot 10^{-3} \eta}\over {D_{kpc}
-R_{kpc}}} \{ {2\over3} B^2 + B_{IC}^2 (1+z)^4 +
7\cdot 10^{-8} {{ L_{QSO}^{46} }\over {R_{kpc}^2}} \}^{-1}
\end{equation}

where $B_{IC}$ is the equivalent magnetic field associated 
to the CMB and  
$L_{QSO}^{46}$ is the luminosity of the
hidden quasar in units of $10^{46}$erg s$^{-1}$.
The AIC scattering with the nuclear photons dominates 
the radiative 
losses at relative small distance from the nucleus.
This distance depends
on the quasar luminosity and on the magnetic field strength
in the lobes (Fig.13).

Due to the decrease of the break energy
with decreasing distance from the nucleus, the innermost regions
of the radio volume contribute less to the radiation spectrum
of the radio galaxies at high frequencies.
Spatially resolved spectroscopy 
in the X--ray domain can provide important information about
$\gamma_b(R)$, especially in the innermost regions of the radio 
galaxies.
The break energy can also be constrained by deep 
radio spectroscopy.
However, it should be noticed that in the innermost parts, 
where $\gamma_b(R) \leq 10^3$,  
the expected repentine decrease of the radio brightness  
makes these investigations very difficult.
If the source is well resolved 
both in the radio and in the X--ray bands, then 
$L_{QSO}$, $B$ and $\eta$ can  
in principle be estimated by comparing the
radio and the X--ray spectral indices with the model predictions.  

In Fig.14,  
for a given radio galaxy model, we have reported
the spectral index 
from the AIC scattering of the nuclear photons 
at different distances from the nucleus.
Fig. 14 is 
obtained by integrating Eq.(31) over the electron distribution
(Eq.44), with $\gamma_b$ given by Eq.(45),  
and over the assumed spectrum of the nuclear photons.
For simplicity, the axis of the 
quasar radiation cone, of half opening angle 45$^o$, 
is assumed to be coincident with that of the radio source
and to lie on the plane of the sky.
In order to give theoretical predictions to be compared
with the future X--ray observations, 
a typical radio galaxy of total size 200 kpc is placed at a 
distance (z=0.15) such that its angular size
is much larger than the Chandra PSF.
Since
the 0.2--10 keV luminosity of a strong FR II radio
galaxy is expected to be close to $\sim 3\cdot 10^{43}$erg s$^{-1}$, 
the estimated Chandra count rate from the 
two regions indicated in Fig.14
is $\sim 10^{-2}$ counts s$^{-1}$
($H_0$=75 km s$^{-1}$ Mpc$^{-1}$, $q_0$=0.0).

The complex spectral index behaviour 
is due to the interplay of 
the IR and UV bumps with the time/spatially evolved particle energy
distribution.
The X--ray spectral index of the innermost regions of the radio galaxies
is significantly steeper than 
those pertaining to the extended regions and to the radio 
synchrotron spectral index of the high brightness lobes.

\begin{figure}
\includegraphics{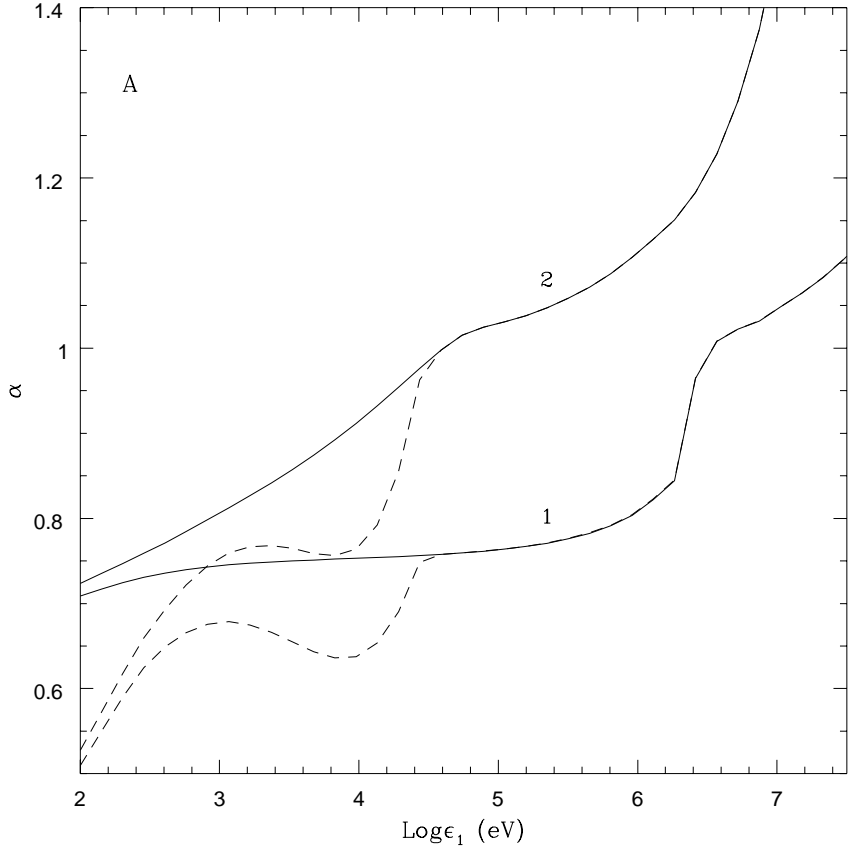}
\vspace{11 cm}
\caption{The calculated spectral index ($j \propto \nu^{-\alpha}$) 
due to the AIC scattering of incoming IR--UV photons
from a typical quasar by a
time--evolved isotropic population of 
relativistic electrons $N(\gamma)\propto \gamma^{-2.5}$
with no energy
cut--off (solid line) and with a low energy cut--off at
$\gamma=20$ (dashed line).
The assumed parameters are those of Fig.13 (case A) with 
the luminosity of the hidden quasar 
$5\cdot 10^{46}$erg s$^{-1}$.
The radio volume is approximated with a 
prolate ellipsoid of $100\times 50$ kpc semiaxis.
The spectra are calculated for two regions comprised between 
2--7 arcsec (2) and 20--25 arcsec (1) from the nucleus ( 
1 arcsec = 2.36 kpc; $H_0$=75 km s$^{-1}$ Mpc$^{-1}$,
$q_0$=0.0).
The synchrotron spectral index of the 
high brightness radio lobes is 0.75.}
\end{figure}

We have also 
computed the X--ray spectral index and emitted power 
by imposing a low energy cut--off in the particle
energy distribution (dashed lines).
In this case the soft X--ray spectral index 
(0.1 -- 1 keV) is expected to be 
considerably harder than the synchrotron spectral index (=0.75).  
As a consequence
detailed spatially resolved spectroscopy 
provided by Chandra may yield unprecedented information 
about the energy distribution of the relativistic
particles in the low energy portion of the spectrum.

If the radio galaxy is inclined with respect to the plane of
the sky the nuclear photons in the far lobe are on average scattered
toward the observer at angles larger than those in the
near lobe.
Since in the case of AIC scattering 
the typical $\gamma$ of the scattering electrons 
depends on the scattering angle (Eq.34), one finds that
in the far lobe the nuclear photons are scattered at a given
$\epsilon_1$ by less energetic electrons
than in the near lobe.
As a consequence, depending on the inclination of the radio galaxy 
on the sky plane, 
all the spectral features in Fig.14 would appear
shifted at lower energies in the case of the near lobe and at larger
energies in the case of the far lobe.
Typically
this results in a spectrum of the far lobe 
significantly harder than that of the near lobe.

\begin{figure}
\includegraphics{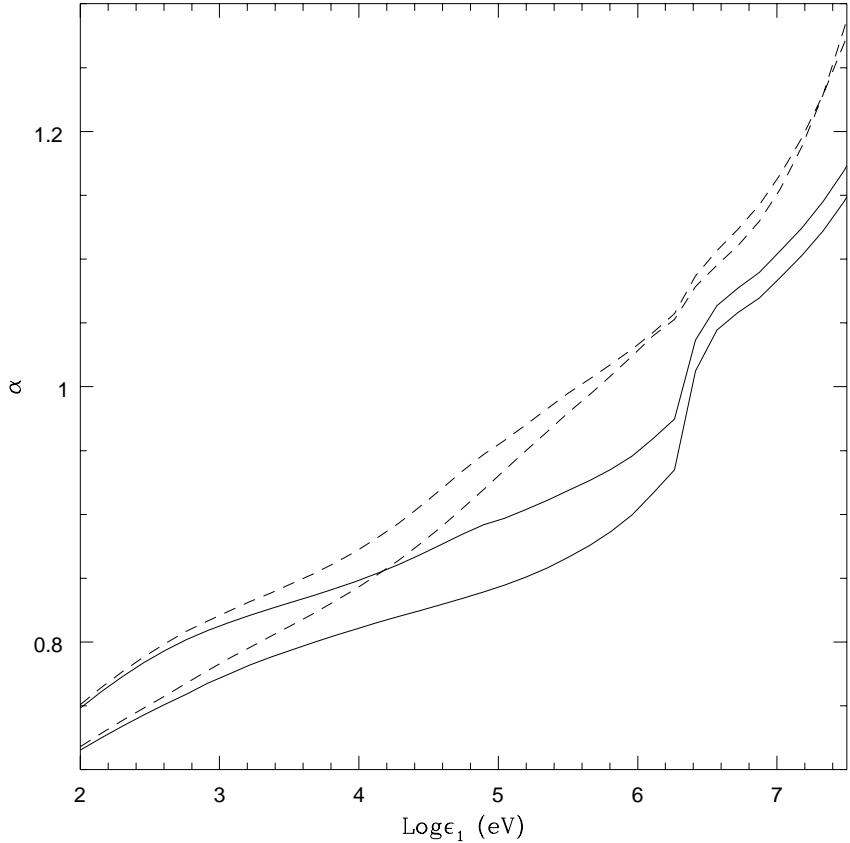}
\vspace{11 cm}
\caption{The calculated spectral index ($j \propto \nu^{-\alpha}$) 
due to the AIC scattering of incoming IR--UV photons
from a typical quasar by an unresolved source.
The assumed magnetic field strength are: 5 (solid lines) and
30$\cdot 10^{-6}$ G (dashed lines).
It is shown for different hidden quasar luminosities,
from the top to the bottom of the diagram one has:
10 and 2 $\cdot 10^{46}$erg s$^{-1}$.
A distance of the hot spot from the nucleus $D$=100 kpc,
a velocity of separation $\eta$=0.02 and a redshift $z$=0.15 have been
assumed.
The radio volume is approximated with a $100\times 50$ kpc prolate
ellipsoid.
The injection index is $\delta$=2.5 so that the radio spectral
index of the high brightness radio lobes is $\alpha$=0.75.}
\end{figure}

\begin{figure}
\includegraphics{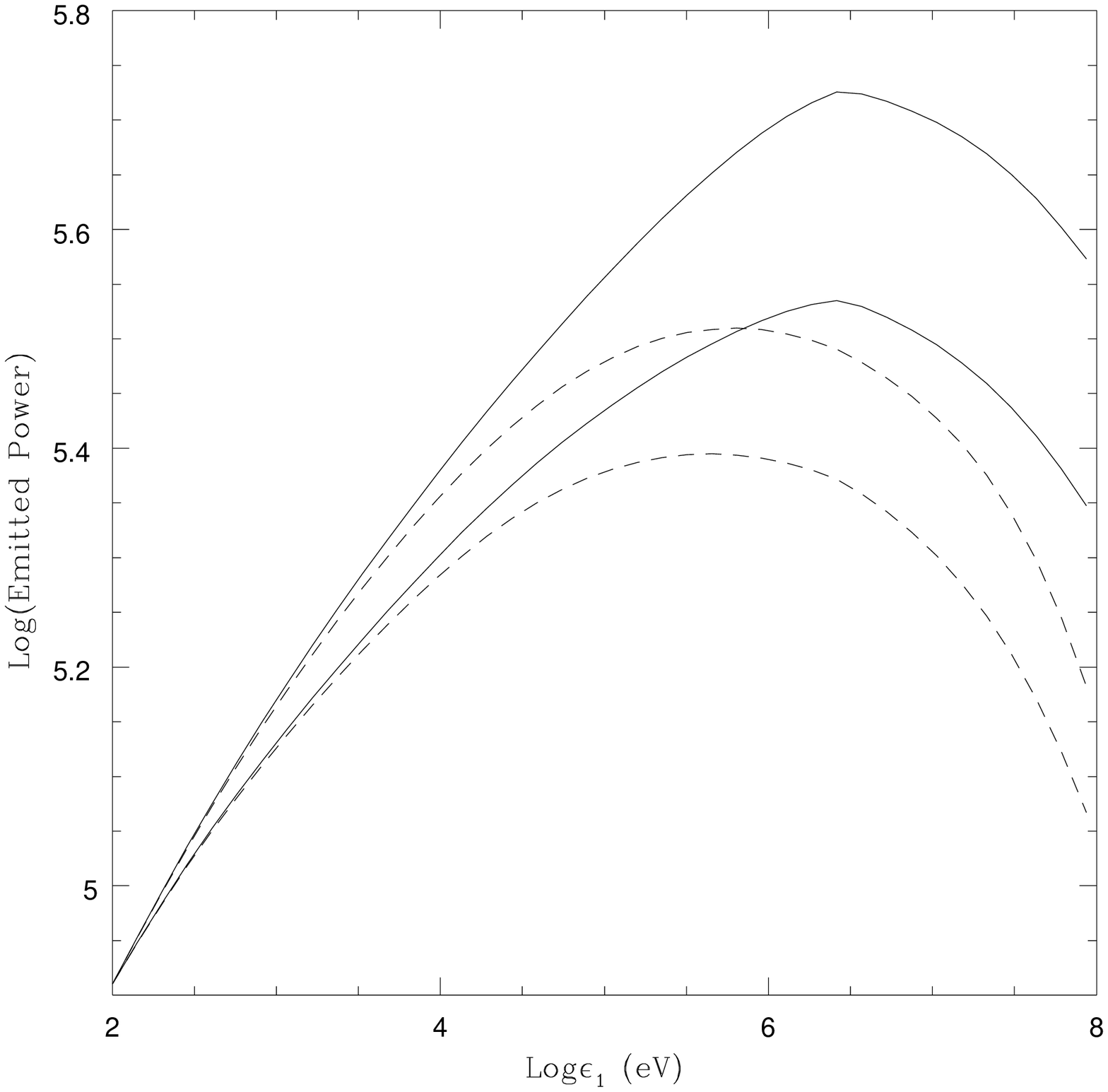}
\vspace{11 cm}
\caption{
The calculated AIC emitted power per unit solid angle
(normalized at 100 eV
and in arbitrary units) due to the scattering of incoming IR--UV photons
from a typical quasar by an unresolved radio galaxy
is shown for different luminosities of the hidden quasar:
2 (top curves) and 10 $\cdot 10^{46}$erg s$^{-1}$ (bottom curves).
The assumed magnetic fields strength are 5 (solid lines) and
30$\cdot 10^{-6}$ G (dashed lines). 
The other parameters are the same of Fig.15.}
\end{figure}

In the case of distant (or faint) radio galaxies spatially resolved 
spectroscopy will not be possible.
However, due to their large effective area and spectral resolution, 
Chandra and XMM could be used to study the integrated spectrum.

The total emitted power is obtained by integrating
the emissivity 
over the emitting volume.
The contribution from the IC scattering of the CMB photons
should be taken into account in the case of
high redshift and/or very extended radio galaxies.
The radio volume can be modelled in detail when one deals with
a specific radio galaxy.
In Figs.14 and 15 the calculated AIC 
spectral index and emitted power (in arbitrary units) from powerful
radio galaxies are shown from the soft X--rays to the gamma--rays for
a number of parameters.
The spectral index from unresolved radio galaxy is
close to the radio spectral index 
in the soft X--ray band, 
being steeper at higher energies.
In the case of a powerful radio galaxy at a redshift
$z$=0.5 with radio spectral index
$\alpha$= 0.75
the expected spectral index in the XMM band would be 
$\alpha_X$= 0.75--0.90 depending on the parameters
(i.e. quasar luminosity, age, magnetic field strength). 
If the radio galaxy is inclined with respect to 
plane of the sky the spectrum from the far lobe would be harder
than that of the near lobe. 
Since the far lobe is also 
expected to be the most luminous (Brunetti et al.1997), 
then the integrated spectrum from an unresolved 
radio galaxy would be harder with
increasing inclination.

From Fig.16 we also notice that 
a large fraction of the energy emitted
by the AIC scattering of the far IR--UV nuclear photons
is channeled in the gamma--ray band.
This will provide a new class of
extragalactic gamma--ray sources to be revealed by future
high energy experiments such as GLAST.

\section{Conclusions}

The anisotropic inverse Compton (AIC) 
scattering problem has been solved from the trans--relativistic to the
ultra--relativistic regime, without 
introducing any approximation.

The equations are integrated over the energies of the scattering
electrons to give the spectrum from the AIC scattering
of an unpolarized monochromatic photon
beam by a beam of unpolarized relativistic electrons.
In the case of a power law energy distribution of the electrons, 
we find that the spectrum is inverted
peaking toward high energies before being sharply cut off
at the emitted energies at which
the highest energetic electrons ($\beta \rightarrow 1$)
are involved in the scattering process.
The emitted power and spectrum in the Thomson approximation is compared
with those in the Klein--Nishina regime.
In the ultra--relativistic Thomson approximation
the radiation scattered in the direction of the electron beam 
has a power law spectrum with slope
$\alpha= (\delta -3)/2$.
We find that  
in the mildly--relativistic case
it is much harder at large scattering angles
and much softer at small scattering angles.

The AIC scattering between 
unpolarized isotropic electrons and a
monochromatic photon beam is discussed from the trans--relativistic 
to the ultra--relativistic case.
In agreement with the previous studies, we find that,
toward the ultra--relativistic limit, the
distribution of the scattered power is highly
anisotropic for the astrophysically common case (where the differential 
electron spectral index $\delta >$ 2) and that the spectrum 
has the same shape of the isotropic ultra--relativistic case.
In addition to previous studies
we find that in the mildly--relativistic case
the spectrum emitted at large scattering angles 
is considerably harder than the classical ultra--relativistic one
being softer at small scattering angles.
Furthermore, we show that the emitted power in the 
Klein--Nishina phase is distributed much more isotropically with 
increasing energy of the scattered photons.
We have also derived simple formulae that give the emissivity from the
ultra--relativistic case down to $\epsilon_1/\epsilon \sim 10$.

In Sect.5 we have applied the general AIC equations
derived in Sect.4 to the calculation of the emitted
power and spectrum from the scattering of the far IR--UV 
nuclear photons from a hidden quasar 
by the relativistic electrons in the lobes of 
powerful FR II radio galaxies.
In our simplified model we have taken into account
the effect of the radiation losses on the
energy distribution of the relativistic electrons.
Due to the losses the calculated spectrum steepens with decreasing
distance from the nucleus where the AIC brightness increases.
This may provide an important test of the model to be performed 
by future X--ray satellites such as Chandra and, conversely, it may
provide extremely important information on the magnetic field strengths
and on the time evolution of the relativistic particles in the FR II 
radio galaxies.
In the case of resolved 
radio galaxies inclined with respect to the plane 
of the sky, due to the anisotropic illumination
from the nuclear photons, we predict that 
the X--ray spectrum of the far lobe is on the average 
harder than that of the near one; since
the far lobe is expected to be the more luminous, 
also the spectrum of unresolved
sources will be harder with increasing inclination.

\begin{ack}
I am indebted to Prof. G.Setti for many discussions
and for suggestions on the manuscript.
I would like to thank Prof. A.Treves for providing useful references
and for useful discussions.
I also thanks the anonymous referee for helpful comments that
have improved the presentation of the paper.
This work was partly supported by the Italian Ministry for
University and Research (MURST) under grant Cofin98-02-32.
\end{ack}

\section{Appendix A: Thomson AIC scattering by isotropic electrons 
with a general energy distribution function}

In Section 4.1 we have derived the Thomson AIC emissivity
by an isotropic electron population 
with energy distribution $f(\gamma)= \gamma^{-\delta}$ (Eqs.35--39).
In several astrophysical situations $f(\gamma)$ may be more
complicated than a simple infinite power law (e.g. Sect.5).
In these cases
the Thomson AIC emissivity is given by: 

\begin{eqnarray}
j(k_3,\epsilon_1) \simeq
{{K_e r_0^2 c}\over{4}} ( {{\epsilon_1}\over{\epsilon_0}} )^2
n
\{
{{2 {\cal I}_0 }\over
{( ( {{\epsilon_1}\over{\epsilon_0}})^2 -2k_3 {{\epsilon_1}\over{\epsilon_0}}
+1)^{1\over 2} }} - \nonumber\\
2(1-k_3)^{2}(1+{{\epsilon_0}\over{\epsilon_1}}) {\cal I}_{3/2}+\nonumber\\
(1-k_3)^{3} 
[(1+{{\epsilon_0}\over{\epsilon_1}})(3k_3-{3\over 2}) -{3\over 2}
({{\epsilon_1}\over{\epsilon_0}}+({{\epsilon_0}\over{\epsilon_1}})^2)]
{\cal I}_{5/2}
\nonumber\\
+{5\over 2}(1-k_3)^{5}
[3(1+{{\epsilon_0}\over{\epsilon_1}})+({{\epsilon_1}\over
{\epsilon_0}}+ ({{\epsilon_0}\over{\epsilon_1}})^2 ) ] 
{\cal I}_{7/2} \}
\end{eqnarray}

with
 
\begin{equation}
{\cal I}_0 = \int_{\gamma_{min}}
{{\gamma^{-2} f(\gamma)}\over{(1-\gamma^{-2})^{1/2}}} d\gamma
\end{equation}

\begin{equation}
{\cal I}_{3/2} = \int_{\gamma_{min}}
{{\gamma^{-1} f(\gamma)}\over{(1-\gamma^{-2})^{1/2}}} 
\{\gamma^2(1-k_3)^2+1-k_3^2 \}^{-3/2}
d\gamma
\end{equation}

\begin{equation}
{\cal I}_{5/2} = \int_{\gamma_{min}}
{{\gamma^{-1} f(\gamma)}\over{(1-\gamma^{-2})^{1/2}}} 
\{\gamma^2(1-k_3)^2+1-k_3^2 \}^{-5/2}
d\gamma
\end{equation}

\begin{equation}
{\cal I}_{7/2} = \int_{\gamma_{min}}
{{\gamma f(\gamma)}\over{(1-\gamma^{-2})^{1/2}}} 
\{\gamma^2(1-k_3)^2+1-k_3^2 \}^{-7/2}
d\gamma
\end{equation}

which, in general, can be readily numerically calculated with
$\gamma_{min}$ given by Eq.(34).

\section{Appendix B: Thomson AIC scattering by isotropic electrons 
with $f(p)=p^{-\delta}$}

It is well known that Fermi acceleration mechanisms lead to an 
energy distribution of the relativistic particles 
that is a power law in momentum and
not in $\gamma$.
As a consequence, in some astrophysical situations in which 
trans and mildly--relativistic AIC scattering produces radiation
at relatively large scattering angles (typically $\geq 50-60^o$)
one should use $f(\gamma)=\beta^{-1} (\beta \gamma)^{-\delta}$.

\begin{figure}
\includegraphics{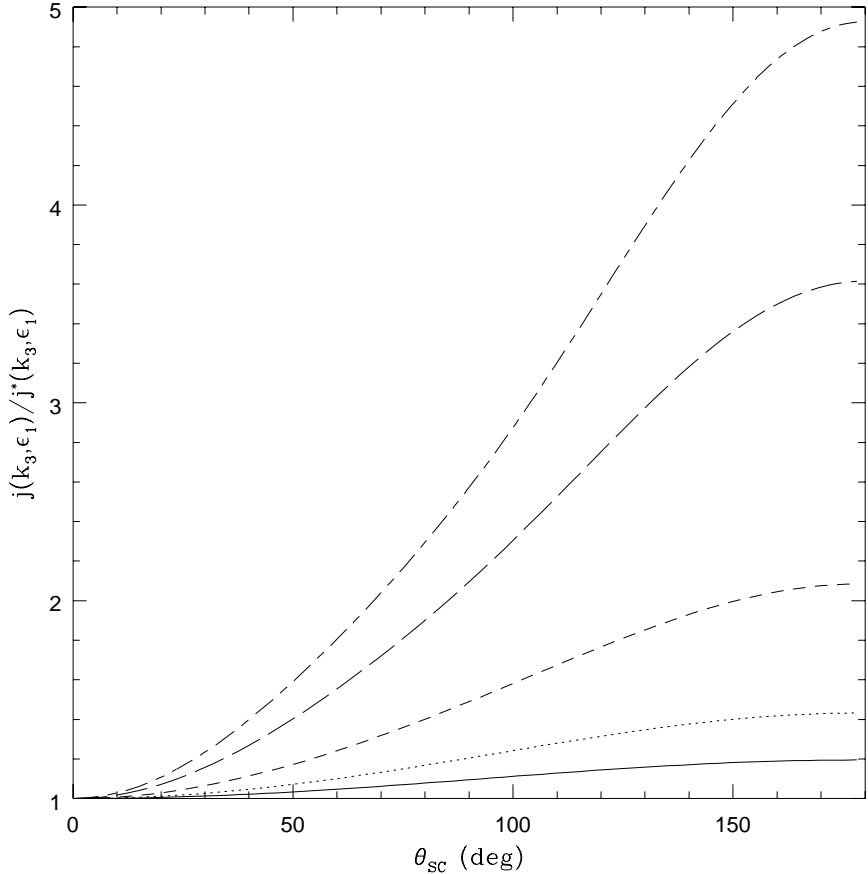}
\vspace{12 cm}
\caption{
The ratio between the angular distribution of the scattered
radiation from Eqs.(35 and 51--54) and from Eqs.(35--39) (i.e.
simple power law in $\gamma$; $j^*$) is shown as a function of the
scattering angle.
In the calculation $\delta$=2.5 has been assumed; the distributions
are given for $\epsilon_1/\epsilon$=16 (solid line), 10 (dotted line),
5 (small--dashed line), and 4 (large--dashed line).
Due to the decreasing energy of the scattering electrons,
the differences between the two calculations are larger
for large scattering angles.}
\end{figure}

In this case the emissivity is calculated from Eqs.(46--50) with 
$f(\gamma)=\beta^{-1} (\beta \gamma)^{-\delta}$. In analitycal form
it is given by Eq.(35) with Eqs.(36--39)
replaced with:

\begin{equation}
{\cal I}_0=
{1\over{\Gamma(\delta/2+1)}} \sum_{i=0}^{\infty}
{{\Gamma(i+1+\delta/2 ) }\over
{(\delta+2i+1)\Gamma(i+1)}}
\gamma_{min}^{-\{\delta+2i+1\} }
\end{equation}

\begin{eqnarray}
{\cal I}_{3/2}=
{{2}\over{\sqrt{\pi}}} \sum_{i,m=0}^{\infty}
{{(m+{1\over 2}) \Gamma(m+{1\over 2}) \Gamma(i+ 1 +\delta/2)
 }\over{
(g_{i,m}-2) \Gamma(i+1) \Gamma(m+1) \Gamma(\delta/2+1)}}
\gamma_{min}^{-\{g_{i,m}-2\} } S^m_{k_3}
\end{eqnarray}

\begin{eqnarray}
{\cal I}_{5/2}=
{4\over{ 3 \sqrt{\pi}}}
\sum_{i,m=0}^{\infty}
{{ (m^2+2m+{3\over 4}) \Gamma(m+{1\over 2}) 
\Gamma(i+ 1 +\delta/2)
 }\over{
g_{i,m} \Gamma(m+1) \Gamma(i+1) \Gamma(\delta/2+1 ) }}
\gamma_{min}^{-g_{i,m} } S^m_{k_3}
\end{eqnarray}

\begin{eqnarray}
{\cal I}_{7/2}=
{8\over{15 \sqrt{\pi}}}
\Gamma(\delta/2+1 )^{-1} \cdot \nonumber\\
\sum_{i,m=0}^{\infty}
{{ (m^3+ {9\over 2} m^2 + {{23}\over 4} m + {{15}\over 8} )
\Gamma(m+{1\over 2}) \Gamma(i+1+\delta/2) 
}\over{g_{i,m} \Gamma(i+1) \Gamma(m+1) }}
\gamma_{min}^{-g_{i,m} } S^m_{k_3}
\end{eqnarray}

where $S_{k_3} \equiv (k_3^2-1)/(1-k_3)^2$, 
$g_{i,m} \equiv \delta+2(i+m)+5$, and $\gamma_{min}$ is given
by Eq.(34).

The differences between the calculation with 
a simple electron energy
distribution $f(\gamma)=\gamma^{-\delta}$ and the correct 
distribution  $f(\gamma)=\beta^{-1} (\beta \gamma)^{-\delta}$ 
are important
for $\epsilon_1/\epsilon < 20$ (Fig.17).

\vskip 1 truecm

{\bf References}

\par\medskip\noindent
Aharonian F.A., Atoyan A.M., 1981, ApSS 79, 321
\par\medskip\noindent
Barthel P. D., 1989, ApJ 336, 606
\par\medskip\noindent
Baylis W.E., Schmid W.M., L\"{u}scher E., 1967, Zei. f\"{u}r
Astrophys. 66, 271
\par\medskip\noindent 
Bednarek W., Kirk J.G., 1995, A\&A 294, 366
\par\medskip\noindent
Bednarek W., 1998, MNRAS 294, 439
\par\medskip\noindent
Berestetskii V. B., Lifshitz E. M., Pitaevski L. P., 1982.{\it
Quantum Electrodynamics, Landau and Lifshitz Course of Theoretical
Physics}, Vol.4, 2nd edn, Pergamon Press, Oxford.
\par\medskip\noindent
Blumenthal G. R., Gould R. J., 1970, Rev. of Mod. Phys. 42, 237
\par\medskip\noindent
Bonometto S., Cazzola P., Saggion A., 1970, A\&A 7, 292 
\par\medskip\noindent
B\"ottcher M., Mause H., Schlickeiser R., 1997, A\&A 324, 395
\par\medskip\noindent
Brunetti G., 1998, PhD Thesis, Dep. of Astronomy, Bologna University
\par\medskip\noindent
Brunetti G., Setti G., Comastri A., 1997, A\&A 325, 898
\par\medskip\noindent
Brunetti G., Comastri A., Setti G., Feretti L., 1999,
A\&A 342, 57
\par\medskip\noindent
Carilli C.L., Perley R.A., Dreher J.W., Leahy J.P.,
1991, ApJ 383, 554
\par\medskip\noindent
Coppi P. S., Blandford R. D., 1990, MNRAS 245, 453
\par\medskip\noindent
Dermer C. D., Schlickeiser R., 1993, ApJ 416, 458
\par\medskip\noindent
Fargion D., Konoplich R.V., Salis A., 1997, Z.Phys.C 74, 571
\par\medskip\noindent
Ghisellini G., George I. M., Fabian A. C., Done C., 1991, 
MNRAS 248, 14
\par\medskip\noindent
Kardashev N.S., 1962, Sov. Astr. 6, 317 
\par\medskip\noindent
Jaffe W.J. Perola G.C., 1973, A\&A 26, 423
\par\medskip\noindent
Jones F. C., 1968, Phys. Rev. 167, 1159
\par\medskip\noindent
Morini M., 1981, ApSS 79, 203
\par\medskip\noindent
Morini M., 1983, MNRAS 202, 495
\par\medskip\noindent
Moskalenko I.V., Strong A.W., 1999 submitted to ApJ; astro-ph/9811284
\par\medskip\noindent
Nagirner D.I., Poutanen J., 1993, A\&A 275, 325 
\par\medskip\noindent
Pozdnyakov L.A., Sobol I.M., Syunyaev R.A., 1983, Sov. Sci.
Rev. E Astrophys. Space Phys. 2, 189
\par\medskip\noindent
Protheroe R.J., Mastichiadis A., Dermer C.D., 1992, Astropart.Phys.
1, 113
\par\medskip\noindent
Rybicki G.B., Lightman A.P., "Radiative Processes in 
Astrophysics", 1979, John Wiley \& Sons Eds.
\par\medskip\noindent
Walter R., Fink H. H., 1993, A\&A 274, 105

\end{document}